\documentclass[graybox]{svmult}

\usepackage{mathptmx}       
\usepackage{helvet}         
\usepackage{courier}        
\usepackage{type1cm}        
\usepackage{makeidx}         
\usepackage{graphicx}        
\usepackage{multicol}        
\usepackage[bottom]{footmisc}

\makeindex             

\begin{document}

\title*{Nonlinear thermoelectric response of quantum dots: renormalized dual fermions out of equilibrium}
\titlerunning{Nonlinear thermoelectric response of quantum dots} 
\author{Stefan Kirchner, Farzaneh Zamani, and Enrique Mu\~{n}oz}
\institute{Stefan Kirchner \at Max Planck Institute for Chemical Physics of Solids, N\"{o}thnitzer Str. 40,  and Max Planck Institute for Physics of Complex Systems, N\"{o}thnitzer Str. 38, 01187 Dresden, Germany, \email{kirchner@pks.mpg.de}
\and Farzaneh Zamani \at Max Planck Institute for Physics of Complex Systems, N\"{o}thnitzer Str. 38, 01187 Dresden, Germany, \email{farzaneh@pks.mpg.de}
\and Enrique Mu\~{n}oz \at Facultad de Fisica, Pontificia Universidad Catolica de Chile, Casilla 306, Santiago 22, Chile, \mbox{\email{ejmunozt@uc.cl}}}

%
%
\maketitle


\abstract{The thermoelectric transport properties of nanostructured devices  continue to attract attention
from theorists and experimentalist alike as the spatial confinement allows for a controlled approach to transport properties of correlated matter. Most of the existing work, however,  focuses on thermoelectric transport in the linear regime despite the fact that the nonlinear conductance of correlated quantum dots has been studied in some detail throughout the last decade.
Here, we review our recent work on the effect of particle-hole asymmetry on the nonlinear transport properties in the
vicinity of the strong coupling limit of Kondo-correlated quantum dots and extend the underlying method, a renormalized superperturbation theory on the Keldysh contour, to the thermal conductance in the nonlinear regime.  We determine the charge, energy, and heat current through the nanostructure and study the nonlinear transport coefficients, the entropy production, and the fate of the Wiedemann-Franz law in the non-thermal steady-state.
Our approach is based on a renormalized perturbation theory in terms of dual fermions around the
particle-hole symmetric strong-coupling limit.}

\section{Introduction}
\label{sec:1}
The ability to transform energy from one form to another is of great socio-economical importance. Electricity plays in this context a special role as modern societies tend to rely on its permanent availability. Yet, the efficiency
with which the energy stored in the chemical bonds of fossil fuels 
is transformed into electricity is only about 30\% while the efficiency at which photovoltaic elements turn the energy of photons into electricity is, at the time of writing, at a level of about 20\%  in commercially available photovoltaic cells. The major part of the stored energy ends up as heat. Utilizing
part of this waste heat for example via the Seebeck effect in a thermoelectric generator is evidently of great practical interest but, as with all heat engines, the efficiency of this process is ultimately limited by that of the ideal Carnot cycle, $\eta_{\mbox{\tiny carnot}}=1-T_{\mbox{\tiny cold}}/T_{\mbox{\tiny hot}}$, where $T_{\mbox{\tiny cold}}/T_{\mbox{\tiny hot}}$ is the temperature of the cold/hot reservoir respectively.
The proportionality factor between the efficiency of the thermoelectric generator and that of the Carnot engine depends on details of charge and heat transfer processes in the heat engine. A quantity of interest is in this context the
dimensionless figure of merit, 
\begin{equation}
Z\bar{T}=\frac{S^2 \sigma \bar{T}}{\kappa},
\label{eqInt1}
\end{equation}
where $\bar{T}=(T_{\mbox{\tiny cold}}+T_{\mbox{\tiny hot}})/2$ is the average temperature,
$S$ is the Seebeck coefficient, $\sigma$ the electrical conductivity, and  $\kappa$ the thermal conductivity.
An increase in the figure of merit results in an enhanced efficiency closer to $\eta_{\mbox{\tiny carnot}}$.
In the limit $Z\bar{T} \rightarrow \infty$; typical  values for $Z\bar{T}$ are of the order of  $Z\bar{T}\approx 1$.

The electrical and thermal conductivity in linear response are defined through
\begin{eqnarray}
\label{eq.:Ls}
{\bf I} &=& L_{11} \nabla V + L_{12} \nabla T,  \\
{\bf Q} &=& L_{21} \nabla V + L_{22} \nabla T. \nonumber
\label{eqInt2}
\end{eqnarray}
where $\bf J$ is the charge current and $\bf Q$ is the heat current through the system in response to
the applied gradients in voltage ($V$) and temperature ($T$) across the sample.
The transport coefficients $L_{i,j}$ are evaluated at equilibrium {\it i.e.} for $\nabla V=0$, $\nabla T=0$ and are not entirely independent, as Onsager's relation requires that $L_{2,1}=TL_{1,2}$~\cite{Onsager.31}.
Onsager's relations ensure that the entropy  production remains semi-positive definite as required by the second law of thermodynamics and are valid beyond the linear response regime.
\newline
The electrical and thermal conductivity are given in terms of $L_{i,j}$ as
\begin{eqnarray}
\sigma &=& L_{11}\\
\kappa &=& \frac{L_{21}L_{12}-L_{22}L_{11}}{L_{11}},
\label{eqInt3}
\end{eqnarray}
and the Seebeck coefficient is defined  by $S=L_{12}/L_{11}$. The definition of $\kappa$ and $S$ reflects that both are defined for vanishing charge current ${\bf I}$.
As the transport coefficients $L_{i,j}$ are evaluated at equilibrium, the fluctuation-dissipation theorem can be invoked to relate the response of the system to its equilibrium fluctuation spectrum~\cite{Callen.51,Kubo.57}.
If the applied gradients in $V$ or $T$
are not sufficiently small, higher order terms
will contribute significantly to $\bf I$ and $\bf Q$ resulting in nonlinear corrections to the electrical and thermal conductivities that require a genuine out-of-equilibrium treatment.
A calculation of the resulting nonlinear conductivities is possible only in certain limiting cases.
\newline
The Boltzmann equation,
\begin{equation}
\frac{\partial F({\bf r},{\bf p})}{\partial t}\Big|_{\mbox{\small collisions}}=
\frac{d{\bf r}}{d t}\nabla_{\bf r} F({\bf r},{\bf p}) 
+ \frac{d{\bf p}}{d t}\nabla_{\bf p} F({\bf r},{\bf p}),
\label{eqInt4}
\end{equation}
{\it e.g.} is a semi-classical equation for the distribution function $F({\bf r},{\bf p})$ in phase space and requires the
existence of well-defined quasi-particles. In addition, further approximations are necessary to evaluate
the collision term. A frequently employed approximation is the relaxation time approximation which assumes that the only effect of the non-equilibrium situation is to
drive the system back to equilibrium. The characteristic rate $\Gamma$, in which the non-equilibrium state decays is then set by the relaxation time $\tau$ ($\Gamma\sim 1/\tau$).
In the relaxation time approximation, the collision term is given by
\begin{equation}
\frac{\partial F({\bf r},{\bf p})}{\partial t}\Big|_{\mbox{\small collisions}}=
-\frac{F({\bf r},{\bf p})-F_0({\bf r},{\bf p})}{\tau},
\label{eqInt5}
\end{equation}
where $F_0$ is the equilibrium distribution function.

For an ordinary metal, well described by Landau's phenomenological Fermi liquid theory, the thermal and charge transport are intimately linked as both are due to the same quasi-particles. This is the content of the 
Wiedemann-Franz law.
This law states that in the limit of purely elastic scattering, the ratio of $\kappa$ and the product of $\sigma$ and $T$ approaches a constant,
\begin{equation}
\lim_{T\rightarrow 0}\frac{\kappa}{\sigma T} = L_0,
\label{eqInt6}
\end{equation}
where $L_0=\pi^2 k_B^2/(3 e^2)$ is the Lorenz number ($k_B$ is Boltzmann's constant and $|e|$ is the charge quantum). It is worth stressing that in general any inelastic scattering, e.g. with phonons or magnons may contribute to the thermal conductivity at any finite $T$: $\kappa=\kappa_{\mbox{\tiny electron}}+\kappa_{\mbox{\tiny phonon}}+\kappa_{\mbox{\tiny magnon}}+\ldots$ but at $T=0$, $\kappa=\kappa_{\mbox{\tiny electron}}$ in a Fermi liquid.
As a consequence of the Wiedemann-Franz law,  the figure of merit, $Z\bar{T}$, of a metal at sufficiently low $T$ is determined by the thermopower (or Seebeck coefficient) $S$ which  is
typically small. The Seebeck coefficient $S$ of a simple metal can be estimated from Mott's formula~\cite{Jonson.80}. 
\newline
One possible route to obtaining higher values of $Z\bar{T}$ in metals is in utilizing regimes where
the Wiedemann-Franz law does not hold. In a superconductor {\it e.g.} one finds $\kappa/(\sigma T)=0$ 
but the thermopower vanishes also since the flow of charge in a superconductor does not give rise to a heat current.
One-dimensional metals violate the Wiedemann-Franz law as well~\cite{Wakeham.11}.
In certain intermetallic rare-earth metals that display  quantum criticality
the Wiedemann-Franz law is also violated~\cite{Tanatar.07,Pfau.12}. 
As the system is quantum critical, the low-lying excitations are scale-invariant and very different from those of a Fermi liquid. As a result, neither the Boltzmann equation is applicable to treat transport due to the absence of well-defined quasi-particles, nor is a linear-response treatment warranted, as no intrinsic scale is present compared to which the applied gradients can be considered small~\cite{Kirchner.09b}. It therefore is to be expected that these systems have a rich
out-of-equilibrium behavior with interesting thermoelectric properties~\cite{Kirchner.10a}.

A particular promising route to relatively  high values of $Z\bar{T}$ has been offered by nanostructured devices and by superlattice structures of correlated materials~\cite{Venkatasubramanian.01,Harman.02,Majumdar.04,Poudel.08,Chowdhury.09,Zhang.11}.
Nanostructured devices also allow for a controlled way of addressing the nonlinear transport regime.
Yet, nonlinear thermal transport properties have so far only received limited attention. This is largely due to the lack of reliable methods which allow for the accurate calculation of nonlinear transport coefficients
in strongly correlated systems.
A noteable expection is some recent work on the nonlinear thermal transport through a molecular junction coupled to local phonons based on rate equations~\cite{Leijnse.10}. Although it remains unclear if this approach does give reliable transport properties at low temperatures, the authors
find strong enhancement of the nonlinear transport coefficients over their linear response counterparts.

An enhancement of the nonlinear thermoelectric transport coefficients over their linear-response counterparts seems natural: 
relaxation processes occurring at finite $T$ and at finite bias voltage do not enter the transport coefficients on equal footing so that 
the breakdown of the Wiedemann-Franz law will as functions of $T$ and at finite non-equilibrium drive may occur differently.
As a result, the nonlinear thermal transport regime may indeed be key in the search for optimal efficiency of thermoelectric heat engines.

Here, we focus on the electronic contribution to the thermoelectric transport properties of strongly correlated quantum dots. In particular, we study the behavior of the heat and charge current through a quantum dot described by
the single-level Anderson model -to be specified below- in the nonlinear transport regime.
We study the nonlinear transport coefficients, the entropy production and the fate of the Wiedemann-Franz law in the nonequilibrium steady-state. In accordance with above arguments, we indeed find that e.g. the
nonlinear thermopower is considerably enhanced above its linear-response counterpart.

The linear response regime of the single-level Anderson model has been studied extensively~\cite{Horvatic.79,Costi.94c,Matveev.99,Dong.02,Nguyen.10,Costi.10}. Especially 
Ref.~\cite{Costi.10} gives a complete discussion of the linear transport properties based on
the numerical renormalization group (NRG) method which is known to give accurate results for quantum impurity models.  The extension of these results to the nonlinear regime is difficult as most methods that are
able to capture the physics of strong electron correlations, like e.g. the Bethe Ansatz~\cite{Andrei.83}, NRG~\cite{Bulla.08} and Quantum Monte Carlo~\cite{Rubtsov.05} are at present largely confined to thermal equilibrium. Self-consistent diagrammatic methods like the non-crossing approximation  and perturbative schemes can be extended to the Keldysh contour to treat the non-equilibrium situation. These methods are conserving, as they respect certain Ward identities~\cite{Baym.61}. There is however no
self-consistent method that captures the correct groundstate of the problem~\cite{Kirchner.04}. Perturbation theory in the Coulomb repulsion $U$ on the quantum dot is in principle possible~\cite{Yamada.75,Zlatic.83}.
This perturbative expansion can be reorganized to deal with the strong coupling problem in terms of renormalized parameters~\cite{Hewson.93}. 
As it turns out, the extension of bare perturbation theory in $U$ to the Keldysh contour suffers from an artificial non-conservation of the charge current away from the particle-hole (p-h) symmetric point~\cite{Hershfield.92}.
We recently proposed a scheme on the Kedysh contour that explicitly respects charge conservation even away from p-h symmetry~\cite{Munoz.11} and that builds on the classical work of Yamada and others~\cite{Yamada.75,Zlatic.83}, on Hewson's renormalized perturbation theory to treat the strong coupling limit, and on Oguri's extension to the p-h symmetric Anderson model out of equilibrium~\cite{Hewson.93,Oguri.01,Hewson.05,Hewson.10}, as well as on a superperturbation theory scheme that utilizes dual fermions~\cite{Rubtsov.08,Hafermann.09}. This method is discussed in detail below.
\newline
Our main purpose here is to analyze the nonlinear thermoelectric transport properties of quantum dots whose
low-energy properties are described by a single-impurity Anderson model, in terms of this current conserving
scheme. We demonstrate that it is possible to have in the nonlinear regime an enhanced Seebeck coefficient and a reduced Wiedemann-Franz ($L/L_0$) ratio as compared to their linear response counterparts.

This chapter is organized as follows. In Section \ref{sec:2}, we discuss the issue of current conservation
and introduce the steady-state distribution function of the spin-degenerate single-level Anderson model
model. Section \ref{sec:3} gives an introduction into the method of ~\cite{Munoz.11}, with more details
in Appendix B, and sections \ref{sec:4}--\ref{sec:6}
discusses the nonlinear electric and thermoelectric transport properties of a Kondo-correlated quantum dot.
Appendix A introduces the nonequilibrium Green functions and the Dyson equation on the Keldysh contour.

\section{Current Conservation and the Steady State Distribution Function}
\label{sec:2}
We are interested in describing the transport properties of a small system, {\it i.e.} a system with a discrete spectrum  
and possibly strong (local) Coulomb repulsion, weakly coupled to a continuum of  itinerant degrees of freedom. Despite its apparent simplicity, this class of models captures
very well the low-energy properties of many nanostructured systems ranging from semi-conductor heterostructures to break-junctions and molecular devices~\cite{Manasreh,Ruitenbeek.03,Natelson.06,Nero.10}.\\
We will concentrate on the single-impurity Anderson model (SIAM) with one local spin-degenerate level at
energy $\epsilon_d$ attached to two leads ($L/R$) which are modeled in terms of non-interacting fermions 
and which can be held at different chemical potentials ($\mu_L$ and $\mu_R$).

\subsection{The single-impurity Anderson model out of equilibrium}

The SIAM Hamiltonian is
\begin{equation}
\label{eq:SIAM}
\hat{H} = \hat{H}_c + \hat{H}_d + \hat{H}_{d-c},
\end{equation}
with
\begin{eqnarray}
\hat{H}_c \!\!\!&&= \sum_{\lambda = L, R}
\sum_{k,\sigma}\epsilon_{k\lambda}
\hat{c}_{k\lambda\sigma}^{\dagger}\hat{c}_{k\lambda\sigma}, \\
\hat{H}_d \!\!\!&&= \sum_{\sigma} E_d \hat{d}_{\sigma}^{\dagger}\hat{d}_{\sigma}
+U\left(\hat{d}_{\uparrow}^{\dagger}\hat{d}_{\uparrow} - \frac{1}{2}\right)\left(\hat{d}_{\downarrow}^{\dagger}\hat{d}_{\downarrow} - \frac{1}{2}\right) - \frac{U}{4},
\nonumber\\
\hat{H}_{d-c} \!\!\!&&= \sum_{\lambda = L,
R}\sum_{k,\sigma}\left(V_{k\lambda}\hat{d}_{\sigma}^{\dagger}\hat{c}_{k\lambda\sigma}
+V_{k\lambda}^{*}\hat{c}_{k\lambda\sigma}^{\dagger}\hat{d}_{\sigma}\right).\nonumber
\label{eq:Hamiltonian}
\end{eqnarray}
Here, $\hat{H}_c$ is the Hamiltonian for electrons in a single conduction band at the metallic leads. $\hat{H}_{d}$ is the Hamiltonian
for localized states in the dot, including the Coulomb interaction, and $\hat{H}_{d-c}$ is the coupling term between the
dot and the leads. 
%
The $\hat{c}_{k\lambda\sigma}$ are fermionic operators representing the creation (annihilation)
of electrons in the conduction band of the  left ($\lambda=L$) or right ($\lambda=R$) metallic lead.
Localized states at the central region (quantum dot or molecule) are represented by the fermionic $\hat{d}_{\sigma}$ operators. 
The coefficients $V_{k\lambda}$ represent a scattering potential which couples the quasi-continuum delocalized states at the leads
with the localized states at the central region. 
The density of states of the leads is given by $\rho_{\lambda}=\sum_{\bf k} \delta(\epsilon_{\lambda, \bf k}-\omega)$ and we will assume that $\rho_{L}(\omega)=\rho_{R}(\omega)=\rho(\omega)$.
For simplicity, we also assume that in what follows  $\rho(\omega)$ is p-h symmetric ($\rho(-\omega)=\rho(\omega)$) and that
p-h symmetry is broken only locally.  
For notational convenience, we  introduce $E_d = \epsilon_d + U/2$, such that the p-h symmetric case $\epsilon_{d} = -U/2$ is simply given by $E_{d} = 0$.

\begin{figure}[b]
\sidecaption
\includegraphics[width=8.5cm]{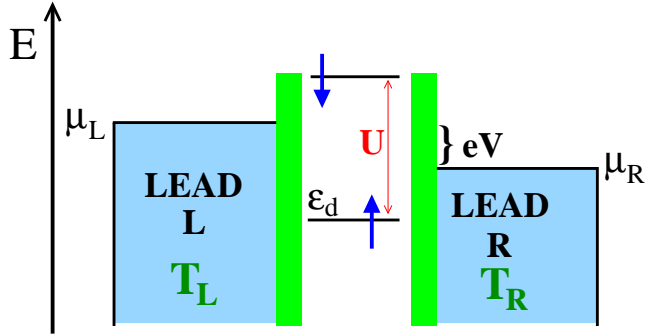}
%
%
\caption{The quantum dot is comprised of a spin-degenerate local level at $\epsilon_d$ and a Coulomb matrix
element $U$. The voltage drop across the quantum dot is set by the difference in chemical potential of the leads, $V=(\mu_L-\mu_R)/e$ and the temperature drop is given by $T_L-T_R$, where $T_L/T_R$ is the temperature in the left/right lead respectively. We choose the zero of energy at the Fermi level of the conducting leads at zero bias voltage.}
\label{fig_dot}       
\end{figure}

Each lead ($\lambda=L/R$) is assumed to be in thermal equilibrium at all times and hence described in terms of an equilibrium distribution function with well-defined  temperature (T$_L$/T$_R$) and chemical potential
($\mu_L/\mu_R$), see Fig.\ref{fig_dot}. The difference in chemical potential ($\mu_L-\mu_R=eV$) and the temperature
difference ($\Delta T=T_L-T_R$) create a particle and energy flux through the central region.
 
Several analytical results are available in the literature
for the p-h symmetric case $E_{d} = 0$, starting with the already classical series
of papers by Yamada and Yosida and others \cite{Yosida.70,Yamada.75,Yamada.76,Yamada.79,Zlatic.83,Horvatic.87}
for the equilibrium case,
and extensions to the non-equilibrium regime by Hershfield and Wilkins \cite{Hershfield.92}, and by Oguri \cite{Oguri.05}. The p-h asymmetric
system, however, has not been studied to the same extent.


The charge current through a nanostructured object attached to non-interacting leads has been
derived in a series of papers. One of the earliest applications of the Keldysh formalism in this context is the calculation of the current through a tunneling junction by Caroli et al.~\cite{Caroli.71}.
A general expression for the charge current through an interacting region in contact with simple  (i.e. non-interacting) leads follows from the continuity equation describing the change in particle number 
in the lead~\cite{Meir.92}.
%
%
%
%
%
%
%
%
%
%
%
%
As shown by Hershfield and Wilkins \cite{Hershfield.92}, the dot
obeys
\begin{eqnarray}
 I_{R} &=& 2\int\frac{d\omega}{2\pi}2\Gamma_{R}[g^{-+}(\omega)[1 - f_{R}(\omega)] - g^{+-}(\omega)f_{R}(\omega)],\nonumber\\
 I_{L} &=& 2\int\frac{d\omega}{2\pi}(-2\Gamma_{L})[g^{-+}(\omega)[1 - f_{L}(\omega)] - g^{+-}(\omega)f_{L}(\omega)],
\label{eqI1}
\end{eqnarray}
where $I_R$ ($I_L$) is the charge current from the right (left) to the dot.
In Eq.~(\ref{eqI1}), we have defined
\begin{eqnarray}
i\Gamma_{\lambda} = -\sum_{k,\sigma}\frac{|V_{k\lambda}|^{2}}{\omega - \epsilon_{k\lambda} + i\eta^{+}}\,\,\,\,\,\,\,\rm{for}\,\,\, \lambda = L, R
\label{eqGamma}
\end{eqnarray}
corresponding to the effective tunneling rate to the metallic leads, so that $\Gamma_{\lambda} \rightarrow \pi\rho_{\lambda}(\omega)|V_{\lambda}|^{2}$ in the limit
of a flat band ($V_{k\lambda} = V_{\lambda}$) of infinite bandwidth, where $\rho_{\lambda}(\omega)=\sum_{k,\sigma}\delta(\omega - \epsilon_{k\lambda})$ is the density of states at the $\lambda = L,\,R$ leads.

It has also been shown in this context that the average of both currents satisfies the relation
\begin{eqnarray}
 \frac{I_{L} + I_{R}}{2} &=& 2\int\frac{d\omega}{2\pi}\frac{2\Gamma_{L}\Gamma_{R}}{\Gamma_{L}+\Gamma_{R}}2\pi A(\omega)
[f_{L}(\omega) - f_{R}(\omega)]\nonumber\\ 
&+& 2\int\frac{d\omega}{2\pi}\frac{1}{2}\frac{\Gamma_{R}-\Gamma_{L}}{\Gamma_{L}+\Gamma_{R}}
[g^{+-}(\omega)\Sigma^{-+}(\omega) - g^{-+}(\omega)\Sigma^{+-}(\omega)],
\label{eqI2}
\end{eqnarray}
whereas the difference, representing the net flux of particles at the central region, is given by
\begin{eqnarray}
 I_{R} - I_{L} = 2\int\frac{d\omega}{2\pi}[g^{+-}(\omega)\Sigma^{-+}(\omega) - g^{-+}(\omega)\Sigma^{+-}(\omega)] = 0.
\label{eqI3}
\end{eqnarray}
In  steady-state, this difference therefore has to vanish,
$I_{R} - I_{L} = 0$. This condition is satisfied, provided
\begin{eqnarray}
 g^{+-}(\omega)\Sigma^{-+}(\omega) - g^{-+}(\omega)\Sigma^{+-}(\omega) = 0.
\label{eqI4}
\end{eqnarray}
This relation certainly holds in equilibrium, where the different components of the self energy and the Green functions
are linked by the Fermi distribution $f_{0}(\omega) = (e^{\hbar\omega/k_{B}T} + 1)^{-1}$,
\begin{eqnarray}
 g^{-+}_{eq}(\omega) &=& 2\pi i A_{eq}(\omega)f_{0}(\omega),\nonumber\\
g^{+-}_{eq}(\omega) &=& 2\pi i A_{eq}(\omega)[1 - f_{0}(\omega)],\nonumber\\
\Sigma^{-+}_{eq}(\omega) &=& \left[ \Sigma^{r}_{eq}(\omega) - \Sigma^{a}_{eq}(\omega)\right]f_{0}(\omega),\nonumber\\
\Sigma^{+-}_{eq}(\omega) &=& \left[ \Sigma^{r}_{eq}(\omega) - \Sigma^{a}_{eq}(\omega)\right][1-f_{0}(\omega)].
\label{eqI5}
\end{eqnarray}
For an interacting system out of equilibrium, as discussed by Hershfield et al.\cite{Hershfield.92}, an effective
distribution function can be defined as follows
\begin{eqnarray}
 g^{-+}(\omega) &=& 2\pi i A(\omega) F_{U}(\omega),\nonumber\\
g^{+-}(\omega) &=& 2\pi i A(\omega)[1 - F_{U}(\omega)].
\label{eqI6}
\end{eqnarray}
Since the Keldysh-Schwinger constraints between the self-energy components are still satisfied for the system out of equilibrium, then in particular we have that
$\Sigma^{+-} + \Sigma^{-+} = \Sigma^{r} - \Sigma^{a}$. Therefore, based on this relation it is possible to define a function $\mathcal{F}(\omega)$ in the following way
\begin{eqnarray}
 \Sigma^{-+}(\omega) &=& \left[ \Sigma^{r}(\omega) - \Sigma^{a}(\omega)\right]\mathcal{F}(\omega),\nonumber\\
\Sigma^{+-}(\omega) &=& \left[ \Sigma^{r}(\omega) - \Sigma^{a}(\omega)\right][1-\mathcal{F}(\omega)].
\label{eqI7}
\end{eqnarray}
Substituting these definitions into Eq.~(\ref{eqI4}), one finds that current conservation in steady-state
is ensured, if
\begin{eqnarray}
A(\omega)\left[ \Sigma^{r}(\omega) - \Sigma^{a}(\omega)\right]\left(\mathcal{F}(\omega)[1 - F_{U}(\omega)]
- F_{U}(\omega)[1 - \mathcal{F}(\omega)] \right)=0.
\label{eqI8}
\end{eqnarray}
 This expression vanishes when $\mathcal{F}(\omega) = F_{U}(\omega)$, that is, in analogy
with the equilibrium situation, in steady-state the Green function components and the self-energy components are related
by the same distribution function $F_{U}(\omega)$. 
As shown in Appendix A, this is indeed the case for the SIAM in the wide-band limit:
\begin{equation}
\label{CurrentConsTest}
\mathcal{F}(\omega) = F_{U}(\omega).
\end{equation}
That the renormalized superperturbation theory does indeed
respect Eq.~(\ref{CurrentConsTest}) and therefore is current conserving was shown in Ref.~\cite{Munoz.11}.

It is instructive to notice that the distribution function for the T-matrix of the interacting SIAM obeys~\cite{Oguri.05}
\begin{eqnarray}
\label{FU}
 F_{U}(\omega) = \frac{\Gamma_{L}f_{L}(\omega) + \Gamma_{R}f_{R}(\omega) + \frac{i}{2}\Sigma^{-+}(\omega)}{\Delta - Im\,\Sigma^{r}(\omega)}.
\end{eqnarray}
Here, we have defined $\Delta=\Gamma_L+\Gamma_R$. If one substitutes the relation $\Sigma^{-+}(\omega) = \left[ \Sigma^{r}(\omega) - \Sigma^{a}(\omega)\right]F_{U}(\omega)$
into Eq.~(\ref{FU}), one finds
\begin{eqnarray}
 F_{U}(\omega) = \frac{\Gamma_{L}f_{L}(\omega) + \Gamma_{R}f_{R}(\omega) -Im\,\Sigma^{r}(\omega)F_{U}(\omega)}{\Delta - Im\,\Sigma^{r}(\omega)}.
\label{eqI10}
\end{eqnarray}
Solving this equation for $F_U(\omega)$ leads to
\begin{eqnarray}
\label{distri}
 F_{U}(\omega) = f_{eff}(\omega) = \frac{\Gamma_{L}f_{L}(\omega) + \Gamma_{R}f_{R}(\omega)}{\Gamma_{L} + \Gamma_{R}}.
\label{eqI11}
\end{eqnarray}
Interestingly, one can arrive at this conclusion from an alternative consideration: The steady-state
condition $I_L-I_R=0$ for the SIAM with identical density of states of left and right leads $\rho_L(\omega)=\rho_R(\omega)=\rho(\omega)$ can be written as
\begin{eqnarray*}
0&=&\frac{i \pi e}{\hbar} \int d\epsilon \rho(\epsilon)\big[
|V|^2_L f_{L} (g^r-g^a)-|V|^2_L F_U(\epsilon)(g^r-g^a)\\
&+& |V|^2_R f_{R} (g^r-g^a)-|V|^2_R F_U(\epsilon)(g^r-g^a)
\big],
\end{eqnarray*}
or
\begin{eqnarray*}
0&=&\int d\epsilon \rho(\epsilon)\big\{ g^r-g^a\big\}\big[
\Gamma_L f_{L}(\epsilon) +\Gamma_R f_{R}(\epsilon)-(\Gamma_L+\Gamma_R) F_U(\epsilon)\big],
\end{eqnarray*}
As $\rho(\omega)$ and $g^r(\omega,T,V)-g^a(\omega,T,V)$ are both semi-positive functions, the steady-state conditions is simply Eq.~(\ref{distri}).

%
Note that the distribution function for the local T-matrix of the SIAM assumes the particularly simple
form of Eq.~(\ref{eqI11}) in the wide-band limit with identical density of states for the left and right lead and in the absence of an external magnetic field. 

\section{Superperturbation theory on the Keldysh contour}
\label{sec:3}
We recently proposed a renormalized non-equilibrium superperturbation theory, in terms of dual fermions
on the Keldysh contour \cite{Munoz.11}. Our primary motivation was to address the issue of current conservation away from p-h symmetry ($E_d\neq0$), Eq.~(\ref{eq:SIAM}).
The term superperturbation theory was introduced in Ref.~\cite{Hafermann.09}, where a quantum impurity coupled to a discrete bath made up of a small  number of bath states was considered as a
reference system. \\
Here, the central idea is to define the interacting ($U\ne 0$) p-h symmetric ($E_{d}=0$) case as a reference system. The solution of the reference system is known  explicitly in terms of a regular expansion in $U$, respectively the renormalized interaction strength $\tilde{u}$~\cite{Yosida.70,Yamada.75,Yamada.76,Yamada.79,Zlatic.83,Horvatic.87,Hewson.93,Oguri.05}.
An expansion around this reference system is expected to work well, as the potential scattering term is
marginally irrelevant.

The retarded local Green function $G_{\sigma,\omega}^{r}=-i\Theta(\tau-\tau')\langle [d^{}_{\sigma}(\tau),d^{\dagger}_{\sigma'}(\tau')] \rangle$ near the strong-coupling fixed point in the presence of p-h asymmetry within renormalized superperturbation theory becomes~\cite{Munoz.11}
\begin{eqnarray}
\label{Gret}
 G_{\sigma,\omega}^{r} &=& (\omega + i\Delta - \Sigma_{E_{d}}^{r})^{-1}\nonumber\\
&=&\tilde{\chi}_{++}^{-1}\left(\omega - \tilde{E}_{d} + i\tilde{\Delta} + \tilde{E}_{d}\tilde{u}\left\{1 - \frac{1}{3}\left[\left(\frac{\pi T}{\tilde{\Delta}} \right)^{2}+\zeta\left(\frac{eV}{\tilde{\Delta}} 
\right)^{2}\right]
+ \frac{2\zeta}{3}\right.\right.\nonumber\\
&&\left.\left.\times\left(\frac{\pi T e V}{\tilde{\Delta}^{2}} \right)^{2}
\right\}
+i\frac{\tilde{\Delta}}{2}\tilde{u}^{2}\left[\left(\frac{\omega}{\tilde{\Delta}} \right)^{2} + \left(\frac{\pi T}{\tilde{\Delta}} \right)^{2}
+ \zeta\left(\frac{eV}{\tilde{\Delta}} \right)^{2} \right]\right)^{-1},
\end{eqnarray}
where the renormalized parameters are given by the expressions:
\begin{eqnarray}
\tilde{u} = \tilde{\chi}_{++}^{-1}U/(\pi\Delta),
\label{eqSPT1}
\end{eqnarray}
which represents the renormalized Coulomb interaction,
\begin{eqnarray}
\tilde{E}_{d} \equiv \tilde{\chi}_{++}^{-1} E_{d}
\label{eqSPT2}
\end{eqnarray}
representing the p-h asymmetry, and
\begin{eqnarray}
 \tilde{\Delta} \equiv \tilde{\chi}_{++}^{-1}\Delta,
\label{eqSPT3}
\end{eqnarray}
being the renormalized width of the quasiparticle resonance.
The renormalization factor for the quasiparticle Green function is given by $\tilde{\chi}_{++}^{-1}$, with the spin susceptibility
given by the result obtained by Yamada and Yosida \cite{Yosida.70,Yamada.75,Yamada.76} 
\begin{eqnarray}
\tilde{\chi}_{++}= 1 + (3 - \pi^{2}/4)\left(U/\pi\Delta\right)^{2}.
\label{eqSPT4}
\end{eqnarray}
The parameter
\begin{eqnarray}
\zeta = 3\frac{\beta}{(1 + \beta)^{2}},
 \label{eqSPT5}
\end{eqnarray}
with $\beta = \Gamma_{L}/\Gamma_{R}$, is a convenient measure of the asymmetry in the coupling to the leads. In particular, for symmetric coupling, $\beta = 1$, one has $\zeta = 3/4$.
A detailed derivation of the renormalized superperturbation theory around the p-h symmetric SIAM is presented in Appendix B.

The local spectral function $A(\omega,T,V) =-(1/\pi)Im\,G_{\sigma,\omega}^{r}$ within our approach is given by the expression
\begin{eqnarray}
A(\omega,T,V) &=&\frac{\tilde{\chi}_{++}^{-1}}{\pi\tilde{\Delta}}\left(1 + \frac{1}{2}\tilde{u}^{2}\left[\left(\frac{\omega}{\tilde{\Delta}} \right)^{2} + \left(\frac{\pi T}{\tilde{\Delta}} \right)^{2}
+ \zeta\left(\frac{eV}{\tilde{\Delta}} \right)^{2} \right]\right)\left\{\left(\frac{\omega}{\tilde{\Delta}} - \tilde{\epsilon}\right.\right.\nonumber\\
&&\left.\left. + \tilde{\epsilon}\tilde{u}
\left\{1 - \frac{1}{3}\left[\left(\frac{\pi T}{\tilde{\Delta}} \right)^{2}+\zeta\left(\frac{eV}{\tilde{\Delta}} \right)^{2}
\right]+ \frac{2\zeta}{3}\left(\frac{\pi T e V}{\tilde{\Delta}^{2}} \right)^{2}\right\}  \right)^{2}\right.\nonumber\\
&&\left.+ \left(1 + \frac{1}{2}\tilde{u}^{2}\left[\left(\frac{\omega}{\tilde{\Delta}} \right)^{2} + \left(\frac{\pi T}{\tilde{\Delta}} \right)^{2}
+ \zeta\left(\frac{eV}{\tilde{\Delta}} \right)^{2} \right]  \right)^{2}
\right\}^{-1},
\label{eqAret}
\end{eqnarray}
and is the basis for calculating charge and energy current through the quantum dot.
Here, we have defined
\begin{eqnarray}
\tilde{\epsilon} \equiv \tilde{E}_{d}/\tilde{\Delta}
\label{eqSPT6}
\end{eqnarray}
as the degree of p-h asymmetry with respect to the width of the resonance level.
The distribution function $F_U(\omega,T,V)$ of the local T-matrix within our scheme is given by
\begin{equation}
F_U=\frac{\Gamma_L f_L+\Gamma_R f_R}{\Gamma_R+\Gamma_L}-\frac{\zeta \pi^2 \tilde{\chi}^2_{+-}}{12}\big( \frac{T eV}{\Delta}\big)^2 f_0^{\prime \prime}(\omega)+ O(U^4),
\label{eqSPT7}
\end{equation}
where $f_0^{\prime \prime}(\omega)$ is the 2nd derivative of the Fermi function with respect to $\omega$.
Furthermore, 
\begin{eqnarray}
F_U \big(\Sigma^r-\Sigma^a\big)&=&-i\Delta \Big(\frac{U}{\pi \Delta}\Big)^2\Big[\Big(\frac{\omega}{ \Delta}\Big)^2+\Big(\frac{\pi T}{ \Delta}\Big)^2+\zeta \Big(\frac{eV}{\Delta}\Big)^2
\Big]f_{eff}(\omega)\nonumber \\
&=&\Sigma^{-+}(\omega,T,V),
\label{eqSPT8}
\end{eqnarray}
establishing that our approach is indeed current conserving~\cite{Munoz.11}.

\section{Electric conductance in the nonlinear regime}
\label{sec:4}
The electric current in steady-state is calculated from the particle current defined in Eq.~(\ref{eqI2}), 
$I_{e} = e I$. The 
electrical conductance for finite bias voltage across the leads, $\mu_{L} - \mu_{R} = e V$, is defined as
\begin{eqnarray}
 G(T,V) = \left.\frac{\partial (e I)}{\partial V}\right|_{\Delta T = 0}.
\label{eqC0}
\end{eqnarray}
Notice that the definition implies the absence of
a temperature difference between the leads, ($T_{L} = T_{R} = T$). This expression
is calculated from Eq.~(\ref{eqI2}) and Eq.~(\ref{eqAret}). For the purpose of comparing
with existing experimental data, it can be written in the form \cite{Munoz.11}
\begin{eqnarray}
 \frac{G(T,V) - G(T,0)}{G_{0}} &=& c_{V}\left(\frac{eV}{\tilde{\Delta}} \right)^{2} - c_{TV}\left(\frac{eV}{\tilde{\Delta}} \right)^{2}\left(
\frac{k_{B}T}{\tilde{\Delta}}
\right)^{2} - c_{V E_{d}}\left(\frac{e V}{\tilde{\Delta}} \right)\nonumber\\
&&+ c_{TV E_{d}}\left(\frac{e V}{\tilde{\Delta}} \right)^{}\left(
\frac{k_{B}T}{\tilde{\Delta}}
\right)^{2}.
\label{eqC1}
\end{eqnarray}
\begin{figure}[t!]
\vspace*{0.9cm}
\includegraphics[width=10cm]{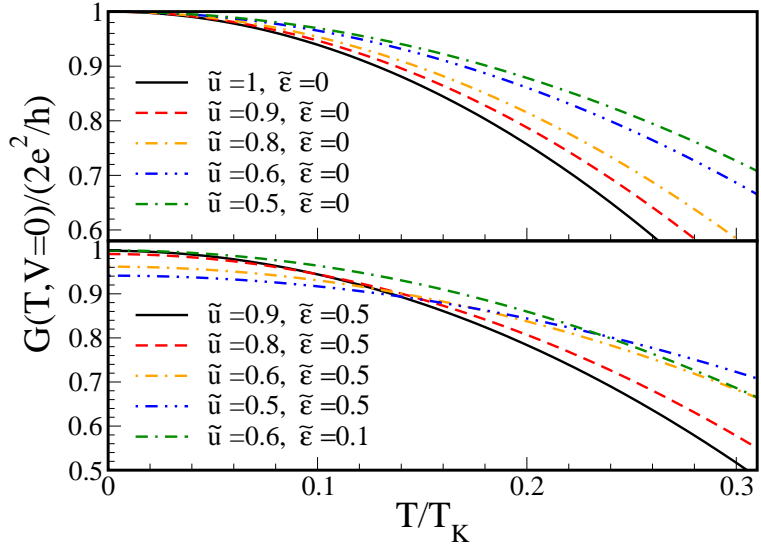}
%
%
\caption{The linear-response conductance, calculated from Eq.~(\ref{eqC3}), in units of twice the quantum of conductance as a function of
temperature for various values of the strength of the renormalized interaction $\tilde{u}$ and p-h asymmetry $\tilde{\epsilon}$. In the presence of p-h asymmetry, the zero-temperature limit of $G(T)$ will be smaller than twice the quantum of conductance in accordance with Friedel's sum rule.
The Kondo temperature $T_K$ here has been defined as $k_B T_K=\pi\tilde{\Delta}/4$.}
\label{conductanceT}       
\end{figure}

The value for the conductance at zero bias voltage and at zero temperature is
\begin{eqnarray}
G_{0} \equiv G(T=0,V=0) = \left(\frac{2 e^{2}}{h} \right)\frac{4}{3}\zeta\left[1 + (1 - \tilde{u})^{2}\tilde{\epsilon}^{2}\right]^{-1},
\label{eqC2}
\end{eqnarray}
with renormalized parameters defined in Eqs. (\ref{eqSPT1}--\ref{eqSPT6}). It is remarkable that this expression satisfies Friedel's sum rule
up to second order in $\tilde{\epsilon}$, $\tilde{u}$, which predicts that the conductance maximum should be 
$G_{0} = (2e^{2}/h)\left[\sin(\pi\langle n_{d} \rangle)\right]^{2} \sim (2e^{2}/h)(1 - \tilde{\epsilon}^{2})$.
The temperature dependence of the electric conductance at zero bias voltage is given by
\begin{eqnarray}
G(T,0) = G_{0}\left[1 - c_{T}\left(\frac{k_{B}T}{\tilde{\Delta}} \right)^{2} \right],
\label{eqC3}
\end{eqnarray}
ans shown in Fig.~\ref{conductanceT} for for different values of $\tilde{\epsilon}$ and $\tilde{u}$.
\begin{figure}[t!]
\phantom{.}
\vspace*{0.9cm}
\includegraphics[width=10cm]{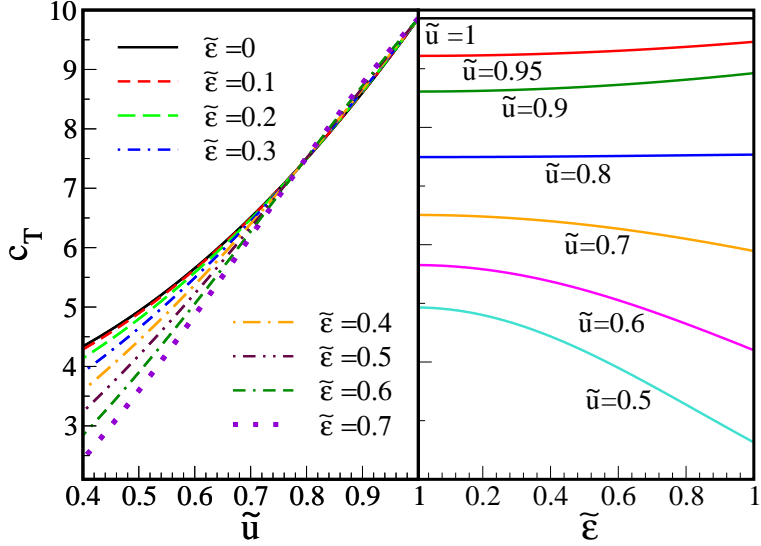}
%
%
\caption{Dependence of $c_T$ on the renormalized quantities $\tilde{u}$ and $\tilde{\epsilon}$ according to Eq.~(\ref{eqC4}). In the strong-coupling limit, {\it i.e.} $\tilde{u}\sim 1$, $c_T$ is in good approximation independent of $\tilde{\epsilon}$ reflecting the marginal irrelevance of the potential scattering term.}
\label{CT}       
\end{figure}
Here, the transport coefficient $c_{T}$ is given by the expression~\cite{Munoz.11}
\begin{eqnarray}
c_{T} = \frac{\pi^{2}}{3}\frac{1 + 2\tilde{u}^2 +\left[\left(8-5 \tilde{u}\right) \tilde{u}-3\right] \tilde{\epsilon}^{2}}{ \left(1 +\left(1-\tilde{u}\right)^{2} \tilde{\epsilon}^{2}\right)^2}.
\label{eqC4}
\end{eqnarray}
Eq.~(\ref{eqC3}) can be compared with a phenomenological formula,
which
is often employed when fitting experimental data in order to obtain the characteristic low-energy ({\it i.e.} Kondo) scale $T_{K}$,
\begin{eqnarray}
 G(T,0) = \frac{G_{0}}{\left(1 + (2^{1/s} - 1)(T/T_{K})^{2} \right)^{s}}.
\label{eqC5}
\end{eqnarray}
Here, $s$ is a phenomenological parameter which is typically taken to be $s\approx 0.2$~\cite{GoldhaberGordon.98}. Note that $c_{T}$ and therefore $s$ is a function of the renormalized interaction strength $\tilde{u}$ and p-h asymmetry $\tilde{\epsilon}$. The variation of $c_T$ with 
 $\tilde{u}$ and $\tilde{\epsilon}$ is shown in Fig.~\ref{CT}.

According to Eqs.~(\ref{eqC3}) and (\ref{eqC5}),
the numerical value of the coefficient $c_{T}$ away from the p-h symmetric point will depend 
on the actual definition used for the Kondo scale $T_{K}$.
The same applies to the remaining transport coefficients of Eq.~(\ref{eqC1}), which are given within our approach by~\cite{Munoz.11}
\begin{eqnarray}
c_{V} &=& \frac{1- \zeta + \tilde{u}^2 (1/2 + \zeta) + \left[\left(\tilde{u}-3\right) \left(\tilde{u}-1\right)\zeta - 3 \tilde{u}\left(\tilde{u}-2\right)-3 \right]\tilde{\epsilon}^{2}}{
\left(1 +\left(1-\tilde{u}\right)^2 \tilde{\epsilon}^{2}\right)^2},\nonumber\\
c_{TV} &=& \pi ^2\left(2 (1 - \zeta) + \tilde{u}^4 (1 + \zeta)
+ 3\tilde{u}^2 (\frac{3}{2} - \zeta) + \left[
\frac{4}{3\pi^{2}}\zeta\tilde{u}(\tilde{u}-1) - 20 + 44\tilde{u}\right.\right.\nonumber\\
&&\left.\left.-\frac{93}{2}\tilde{u}^{2}+46\tilde{u}^{3}-\frac{49}{2}\tilde{u}^{4}
+2\tilde{u}^{5}-\tilde{u}^{6}+(20 - 40\tilde{u} + 35\tilde{u}^{2} - 24\tilde{u}^{3}\right.\right.\nonumber\\
&&\left.\left.+ 8\tilde{u}^{4} + 2\tilde{u}^{5}-\tilde{u}^{6})\zeta\right]\tilde{\epsilon}^{2}
\right)
\left(1+\left(1-\tilde{u}\right)^2 \tilde{\epsilon}^{2}\right)^{-4},\nonumber\\
c_{VE_{d}} &=& 2\left(\frac{1-\beta}{1+\beta}\right)\frac{ (1-\tilde{u})\tilde{\epsilon}}{1+ 
(1-\tilde{u})^2 \tilde{\epsilon}^{2}},\nonumber\\
c_{TV E_{d}} &=& \frac{2 \pi^{2}}{3}\left(\frac{1-\beta}{1 +\beta}\right)\tilde{\epsilon}\left[2
   \left(\tilde{u}-1\right)^2 \left(\tilde{u} \left(3 \left(\tilde{u}-1\right)
   \tilde{u}-4\right)+3\right) \tilde{\epsilon}^2+9 \tilde{u}^3\right.\nonumber\\
&&\left.-9 \tilde{u}^2+6
   \tilde{u}-6\right]\left(1 + \left(1 - \tilde{u}\right)^2 \tilde{\epsilon}^{2}\right)^{-3}.
   \label{eqC6}
\end{eqnarray}
The analytical expression obtained in Eq.~(\ref{eqC1}) can be compared with the 'universal' equation
which has been applied to analyze experimental measurements \cite{Grobis.08,Scott.09} of electrical conductance under steady-state conditions,
for semiconductor heterostructures (quantum dots) and single-molecule devices beyond linear response
\begin{eqnarray}
 \frac{G_{0} - G(T,V)}{c_{T}G_{0}} = \left(\frac{T}{T_{K}} \right)^{2} + \alpha\left(\frac{eV}{k_{B}T_{K}} \right)^{2}
-\gamma c_{T}\left(\frac{e V T}{k_{B}T_{K}^{2}} \right)^{2}.
\label{eqC7}
\end{eqnarray}
Despite the apparently 'universal' form of Eq.~(\ref{eqC7}), different experimental systems
seem to differ in the numerical values of the coefficients $\alpha$ and $\gamma$. In particular,
experiments in GaAs quantum dots \cite{Grobis.08} reported average values of $\alpha_{G} = 0.1$ and $\gamma_{G} = 0.5$,
whereas for single-molecule devices \cite{Scott.09},
considerably smaller values of $\alpha_{S} = 0.01$ and $\gamma_{S} = 0.1$ were obtained.
According to Eqs.(\ref{eqC1}) and Eq.~(\ref{eqC7}), it is clear that the $T_{K}$-independent coefficients $\alpha$ and $\gamma$
can be expressed in terms of the transport coefficients in Eq.~(\ref{eqC6}) by
\begin{eqnarray}
 \alpha = c_{V}/c_{T}, && \gamma = c_{TV}/c_{T}^{2}.
\label{eqC8}
\end{eqnarray}
It is clear from the analytical expressions, Eq.~(\ref{eqC6}), that the numerical values of 
these coefficients are expected to depend on specific sample features, particularly the degree of p-h asymmetry 
$\tilde{\epsilon}$, as well as on the renormalized Coulomb interaction $\tilde{u}$. It is particularly noteworthy that,
in agreement with Fermi liquid theory, in the strongly interacting (Kondo) limit $\tilde{u}\rightarrow 1$,
the transport coefficients in Eq.~(\ref{eqC6}) become independent of the degree of p-h asymmetry 
$\tilde{\epsilon}$.

Our analytical results Eq.~(\ref{eqC6}) explain the numerical values obtained for the transport coefficients
in quantum dot experiments \cite{Grobis.08}, where for instance the set of parameters $\tilde{u} = 0.45$, $\beta = 1$ and
$\tilde{\epsilon}=0.1$ yield $\alpha = 0.1$ and $\gamma = 0.47$, in good agreement with Ref.\cite{Grobis.08}.
On the other hand, our theory cannot explain the particular combination of values for the transport
coefficients in single-molecule experiments \cite{Scott.09}, suggesting that other mechanisms
not captured by the SIAM may play a role in those systems, such as
scattering with local phonons.

\section{Energy transport through the quantum dot and the steady-state entropy production rate}
\label{sec:5}

So far, we have discussed the charge transport through the quantum dot. The charge current is
well defined even in the nonlinear regime  due to charge conservation:
$ \frac{\partial \rho}{\partial t}+\nabla \cdot I=0$,
where $\rho$ is the local charge density and $I$ the associated charge current.
In the present geometry, the continuity equation assumes a particularly simple form
\begin{eqnarray}
 \partial_{t}\langle N \rangle + I_{R} -I_{L} = 0,
\label{eqE0}
\end{eqnarray}
where $\partial_{t}$ represents the partial derivative with respect to time and  $\langle N \rangle$ represents the average local occupation at the dot site.
Clearly, the condition for steady-state is $I_{L} = I_{R} = I$.
The energy current can be introduced in an analogous manner since it is also related to a conserved quantity.
The local energy balance at the spatially localized region, {\it i.e.} the quantum dot, becomes
\begin{eqnarray}
 \partial_{t}\langle E \rangle = J_{E}^{L} - J_{E}^{R}.
\label{eqE1}
\end{eqnarray}
Here, $\langle E \rangle$ represents the average local internal energy,
whereas $J_{E}^{L}$, $J_{E}^{R}$ are the energy currents 
flowing from the left lead to the quantum dot (L), or from the quantum dot to the right lead (R), respectively.

From a similar analysis as for the particle current,
and taking into account that the flow of each quasiparticle involves transport of an
energy quanta $\hbar\omega$,
we have that the net energy currents are given by
\begin{eqnarray}
 J_{E}^{R} &=& 2\int\frac{d\omega}{2\pi}\hbar\omega(2\Gamma_{R})\frac{1}{2i}\left\{g^{K}(\omega) - \left[g^{a}(\omega) - g^{r}(\omega) \right]f^{K}_{R}(\omega) \right\},\nonumber\\
J_{E}^{L} &=& 2\int\frac{d\omega}{2\pi}\hbar\omega(-2\Gamma_{L})\frac{1}{2i}\left\{g^{K}(\omega) - \left[g^{a}(\omega) - g^{r}(\omega) \right]f^{K}_{L}(\omega) \right\}.
\label{eqE12}
\end{eqnarray}
Here, we have defined $f^{K}_{L(R)}(\omega)\equiv 2f_{L(R)}(\omega) - 1$ as the distribution function for each lead. 
It is convenient then to
split these functions in two pieces as follows
\begin{eqnarray}
 f^{K}_{R} &=& f^{K}_{eff} + \frac{\Gamma_{L}}{\Gamma_{L} + \Gamma_{R}}\left(f^{K}_{R} - f^{K}_{L} \right),\nonumber\\
f^{K}_{L} &=& f^{K}_{eff} + \frac{\Gamma_{R}}{\Gamma_{L} + \Gamma_{R}}\left(f^{K}_{L} - f^{K}_{R} \right),
\label{eqE13}
\end{eqnarray}
where we defined $f^{K}_{eff}(\omega) = 2f_{eff}(\omega) - 1$ as the distribution function at the 
local site.
Applying the identity
\begin{eqnarray}
 g^{K}(\omega) - \left[g^{a}(\omega) - g^{r}(\omega) \right]f_{eff}^{K}(\omega) &=& \frac{2i}{2\left(\Gamma_{L} + \Gamma_{R}\right)}\left[g^{+-}(\omega)\Sigma^{-+}(\omega)
\right.\nonumber\\
&&\left.- g^{-+}(\omega)\Sigma^{+-}(\omega) \right],
\label{eqE14}
\end{eqnarray}
one obtains
\begin{eqnarray}
 J^{R}_{E} &=& 2\int\frac{d\omega}{2\pi}\hbar\omega\frac{\Gamma_{R}}{\Gamma_{L} + \Gamma_{R}}\left[g^{+-}(\omega)\Sigma^{-+}(\omega)
- g^{-+}(\omega)\Sigma^{+-}(\omega) \right]\nonumber\\ 
&&+ 2\int\frac{d\omega}{2\pi}\hbar\omega\frac{2\Gamma_{L}\Gamma_{R}}{\Gamma_{L} + \Gamma_{R}}i\left[g^{r}(\omega)
- g^{a}(\omega) \right](f_{L}(\omega) - f_{R}(\omega)),\nonumber\\
J^{L}_{E} &=& 2\int\frac{d\omega}{2\pi}\hbar\omega\frac{-\Gamma_{L}}{\Gamma_{L} + \Gamma_{R}}\left[g^{+-}(\omega)\Sigma^{-+}(\omega)
- g^{-+}(\omega)\Sigma^{+-}(\omega) \right]\nonumber\\ 
&&+ 2\int\frac{d\omega}{2\pi}\hbar\omega\frac{2\Gamma_{L}\Gamma_{R}}{\Gamma_{L} + \Gamma_{R}}i\left[g^{r}(\omega)
- g^{a}(\omega) \right](f_{L}(\omega) - f_{R}(\omega)).
\label{eqE15}
\end{eqnarray}
To check the condition for steady-state in the total energy flow, we substract both currents to obtain
\begin{eqnarray}
J_{E}^{R} - J_{E}^{L} = 2\int\frac{d\omega}{2\pi}\hbar\omega\left[g^{+-}(\omega)\Sigma^{-+}(\omega)
- g^{-+}(\omega)\Sigma^{+-}(\omega) \right].
\label{eqE16}
\end{eqnarray}
It is important to notice that the same condition that we invoked for steady-state in particle flow, {\it i.e.}
\begin{eqnarray}
 g^{+-}(\omega)\Sigma^{-+}(\omega) - g^{-+}(\omega)\Sigma^{+-}(\omega) = 0,
\label{eqE17}
\end{eqnarray}
indeed will also imply energy conservation in steady-state, $J_{E}^{R} - J_{E}^{L} = 0$.
As the renormalized superperturbation is respecting Eq.~(\ref{eqE17}), it is an appropriate tool to study the nonlinear thermoelectric transport properties in a controlled fashion.
 
The steady-state energy 
current $J_{E} = J_{E}^{R} = J_{E}^{L}$ through the quantum dot is finally given by
\begin{eqnarray}
 J_{E} = \frac{J_{E}^{R} + J_{E}^{L}}{2} = \int d\omega \frac{4\Gamma_{L}\Gamma_{R}}{\Gamma_{L}+\Gamma_{R}} \hbar \omega A(\omega)
[f_{L}(\omega) - f_{R}(\omega)].
\label{eqE18}
\end{eqnarray}

For the  generalization of transport coefficients of Eq.~(\ref{eq.:Ls}) other than $L_{11}$, the knowledge of the nonlinear heat current is required. The notion of heat current in spatially extended systems
away from equilibrium is still a matter of debate~\cite{Wu.09,Dubi.11}. In the present case, none of these
difficulties are pertinent as the setup is easily cast into a hydrodynamic language without any approximations.

In the hydrodynamic regime, where the out-of-equilibrium dynamics is only due to low frequency and long wavelength excitations, the system is characterized by a few so-called slow variables that are (away from criticality and in the absence of any Goldstone bosons) are determined entirely through conservation laws.
The resulting {\it local} equilibrium allows for consistent determination of the entropy current $J_S$ via
\begin{equation}
\label{entropycurrent}
 \frac{\partial S^{}}{\partial t}+\nabla \cdot {J_S^{}}={\mathcal{P}},
\end{equation}
where ${\mathcal{P}}>0 $ is the entropy production rate. 
Within the hydrodynamic approach, $\mathcal{P}$ is decomposed into the currents $J_i$  associated with the $N$ conserved quantities:  
\begin{equation}
\label{eq:P}
\mathcal{P}=\sum_{i}^{N} X_i J_i.
\end{equation}
 The currents $J_i$ can also be expressed in terms of the generalized forces $X_i$: 
\begin{equation}
\label{generalizedforces}
 J_i=-\sum_{j} L_{ij} X_j,
\end{equation}
which is nothing but  Eq.~(\ref{eq.:Ls}) and the requirement ${\mathcal{P}}>0 $ is ensured by the Onsager relations~\cite{Onsager.31}.

In the case considered here, where a spatially confined region is attached to non-interaction leads in
thermal equilibrium, we can obtain Eqs.~(\ref{entropycurrent}) and (\ref{generalizedforces}) without resorting to the hydrodynamic limit (but confined to the steady-state) and hence not only obtain the entropy production rate but also determine the transport coefficients $L_{ij}$ in the nonlinear regime.

%
As each lead is characterized by an equilibrium distribution with well-defined
temperature ($T_{L}$/$T_{R}$) and chemical potential ($\mu_{L}$/$\mu_{R}$), the corresponding entropy production rate (${\mathcal{P}}_L$/${\mathcal{P}}_R$) vanishes.
Therefore, the entropy currents from the left lead to the dot, and from the dot to the right lead,
are given by the expressions
\begin{eqnarray}
 T_{L}J_{S}^{L} &=& J_{E}^{L} - \mu_{L}I_{L},\nonumber\\
T_{R}J_{S}^{R} &=& J_{E}^{R} - \mu_{R}I_{R}.
\label{eqE2}
\end{eqnarray}
In steady-state, $\partial_{t}\langle N \rangle = 0$ and
$\partial_{t}\langle E \rangle = 0$ and consequently the energy and particle currents  satisfy
\begin{eqnarray}
J_{E}^{L} &=& J_{E}^{R} = J_{E},\nonumber\\
I_{L} &=& I_{R} = I.
\label{eqE3}
\end{eqnarray}
According to Eq.~(\ref{eqE2}),  the entropy fluxes in steady-state must therefore obey
\begin{eqnarray}
J_{S}^{L} &=& \frac{J_{E}}{T_{L}} - \frac{\mu_{L}}{T_{L}}I,\nonumber\\
J_{S}^{R} &=& \frac{J_{E}}{T_{R}} - \frac{\mu_{R}}{T_{R}}I,
\label{eqE4}
\end{eqnarray}
so that Eq.~(\ref{entropycurrent}) in the present case reads
\begin{eqnarray}
\partial_{t}\langle S \rangle + J_{S}^{R} - J_{S}^{L} = {\mathcal{P}}.
 \label{eqE5}
\end{eqnarray}
Therefore, in  steady-state, where explicit time-dependencies vanish,  $\partial_{t}\langle S \rangle = 0$, and the entropy production 
rate at the dot is found to be
\begin{eqnarray}
{\mathcal{P}} = -J_{E}\Delta\left(\frac{1}{T} \right)
+ I\Delta\left(\frac{\mu}{T}\right),
\label{eqE6}
\end{eqnarray}
where for notational convenience we have defined $\Delta\psi \equiv \psi_{L} - \psi_{R}$.
Eq.~(\ref{eqE6}) together with Eq.~(\ref{eq:P}) allows to identify the generalized forces $X_i$ in the present case.
From Eq.~(\ref{eqE6}) it follows that even under conditions where the charge current vanishes
($I = 0$), there will be  entropy generation at the local dot site (for $T_L \neq T_R$),
\begin{eqnarray}
{\mathcal{P}} = -J_{E}\Delta\left(\frac{1}{T} \right) > 0,
\label{eqE7}
\end{eqnarray}
thus reflecting the existence of an intrinsic dissipation mechanism in order to sustain the steady-state regime.
We notice that after Eq.~(\ref{eqE2}), it is possible to define the heat currents
\begin{eqnarray}
 J_{Q}^{L} &=& T_{L}^{ }J_{S}^{L} = J_{E}^{L} - \mu_{L}I_{L},\nonumber\\
 J_{E}^{R} &=& T_{R}^{ }J_{S}^{R} = J_{E}^{R} - \mu_{R}I_{R},
\label{eqE8}
\end{eqnarray}
where $J_{Q}^{L(R)}$ is identified as a heat current from the left (L) lead to the quantum dot, or from
the quantum dot into the right lead (R).
In steady-state, we have that the heat currents are:
\begin{eqnarray}
 J_{Q}^{L} &=& J_{E} - \mu_{L}I,\nonumber\\
J_{Q}^{R} &=& J_{E}  - \mu_{R}I.
\label{eqE9} 
\end{eqnarray}
Notice that in general $J_{Q}^{L} \ne J_{Q}^{R}$ (total internal energy is conserved, not just heat). 
Moreover, under steady-state conditions ($I_{L} = I_{R} = I$, $J_{E}^{L} = J_{E}^{R} = J_{E}$), 
substitution of 
Eq.~(\ref{eqE9}) into Eq.~(\ref{eqE1})
yields
\begin{eqnarray}
  \partial_{t} \langle E \rangle = J_{Q}^{L} - J_{Q}^{R} + \left(\mu_{L} - \mu_{R} \right)I = 0.
\label{eqE10}
\end{eqnarray}
The corresponding expressions for the heat currents in steady-state are
\begin{eqnarray}
 J_{Q}^{L(R)} = J_{E} - \mu_{L(R)}I = \int d\omega \frac{4\Gamma_{L}\Gamma_{R}}
{\Gamma_{L}+\Gamma_{R}} \left(\hbar \omega -\mu_{L(R)}\right)A(\omega)
[f_{L}(\omega) - f_{R}(\omega)].
\label{eqE19}
\end{eqnarray}

At finite voltage, when $\mu_{L} - \mu_{R} = eV > 0$, the second term in Eq.~(\ref{eqE10}) represents
the macroscopic electric work to sustain the current through the voltage
difference imposed, while the first is the net flow of heat 
at the local site, which is connected with entropy production and dissipation, as previously discussed.
Moreover, Eq.~(\ref{eqE6}) for the local entropy production in steady-state can alternatively be expressed
as
\begin{eqnarray}
 {\mathcal{P}} = \frac{J_{Q}^{R}}{T_{R}} - \frac{J_{Q}^{L}}{T_{L}},
\label{eqE11}
\end{eqnarray}
which is just stating that the local entropy production at the local dot site
must be given by the difference between the rate of entropy gain at the right lead $J_{Q}^{R}/T_{R}$, 
and the entropy loss at the left lead, $J_{Q}^{L}/T_{L}$. 


\section{Thermoelectric transport at finite bias voltage}
\label{sec:6}

Having identified the entropy production rate, the generalized forces and the heat currents, we are now in a position to address the nonlinear generalizations of $L_{12}$ and $L_{22}$ of Eq.~(\ref{eq.:Ls}).

Thermal conductance is experimentally measured under conditions such
that the electric current vanishes. This leads to the
fairly general definition
\begin{eqnarray}
 K(T,V) = \left.\frac{\partial J_{Q}}{\partial (\Delta T)}\right|_{I=0},
\label{eqK1}
\end{eqnarray}
which is valid regardless of the thermal gradients and bias voltages
being infinitesimal or finite, and therefore is applicable beyond
the linear response regime. In the previous section, we obtained
expressions for the heat currents in steady-state conditions, Eq.~(\ref{eqE19}),
$J_{Q}^{L(R)} = J_{E} - \mu_{L(R)}I$. In particular, if we restrict ourselves to
the condition of a vanishing charge current ($I=0$), we have that the heat currents satisfy $J_{Q}^{L}|_{I=0} = J_{Q}^{R}|_{I=0}$,
with 
\begin{eqnarray}
 \left.J_{Q}\right|_{I=0} = J_{E} = \int d\omega 
\frac{4\Gamma_{L}\Gamma_{R}}{\Gamma_{L}+\Gamma_{R}} \hbar \omega A(\omega,T,V) [f_{L} - f_{R}],
\label{eqK2}
\end{eqnarray}
given by Eq.~(\ref{eqE19}).
Therefore, the problem of calculating the thermal conductance can be stated as
\begin{eqnarray}
 K(T,V) = \frac{\partial}{\partial(\Delta T)}\left(\int d\omega \frac{4\Gamma_{L}\Gamma_{R}}{\Gamma_{L}+\Gamma_{R}} 
\hbar \omega A(\omega,T,V) [f_{L} - f_{R}] \right)_{I=0},
\label{eqK3}
\end{eqnarray}
subject to the condition
\begin{eqnarray}
 I = \int d\omega \frac{4\Gamma_{L}\Gamma_{R}}{\Gamma_{L}+\Gamma_{R}} A(\omega,T,V)[f_{L} - f_{R}] = 0.
\label{eqK4}
\end{eqnarray}
It is clear that the condition of vanishing particle current Eq.~(\ref{eqK4}) is fulfilled
when the thermal gradient and the bias voltage are related. This
relation is explicitly given by the definition of the Seebeck coefficient $S(T,V)$,
\begin{eqnarray}
 S(T,V) = \left.\frac{\partial V}{\partial (\Delta T)}\right|_{I = 0},
\label{eqK5}
\end{eqnarray}
which will be discussed in detail in the next section.
It is more convenient to express Eq.~(\ref{eqK3}) via the implicit
function differentiation rule. Since Eq.~(\ref{eqK4}) defines an implicit functional
relation $V = V(\Delta T)|_{I=0}$, one has
\begin{eqnarray}
 \left.\frac{\partial \psi(\Delta T, V(\Delta T))}{\partial \Delta T}\right|_{I=0} 
= \left.\frac{\partial \psi}{\partial \Delta T}\right|_{V}
+ \left.\frac{\partial V}{\partial \Delta T}\right|_{I=0}\left.\frac{\partial \psi}{\partial V}\right|_{\Delta T}.
\label{eqK6}
\end{eqnarray}
Applied to the thermal conductance, Eq.~(\ref{eqK3}) becomes
\begin{eqnarray}
 K(T,V) &=& \int d\omega \frac{4\Gamma_{L}\Gamma_{R}}{\Gamma_{L}+\Gamma_{R}} 
\hbar \omega A(\omega,T,V) \frac{\partial}{\partial(\Delta T)}[f_{L} - f_{R}]\nonumber\\
&&+S(T,V)\frac{\partial}{\partial V}\left(
\int d\omega \frac{4\Gamma_{L}\Gamma_{R}}{\Gamma_{L}+\Gamma_{R}} 
\hbar \omega A(\omega,T,V)[f_{L} - f_{R}]
\right)_{\Delta T}.
\label{eqK7}
\end{eqnarray}

In the linear response regime, the situation is relatively simple, since
it is sufficient to 
expand $f_{L} - f_{R} = -(\Delta T/T)\omega\partial f_{0}/\partial\omega + (eV)\partial f_{0}/\partial\omega$, 
and to substitute $A(\omega,T,V=0)$ in the integrand of Eqs.(\ref{eqK4}) and (\ref{eqK5}). As a result,
the relation between temperature gradient and voltage is linear, and given by
$\left.\frac{\partial V}{\partial (\Delta T)}\right|_{I = 0} = S(T,0)$ in Eq.~(\ref{eqK5}).

In the nonlinear regime, however, the relation is not trivial at all, since the nonlinear
relation between bias voltage and temperature gradient which satisfies the zero electric current
condition is implicitly given
by Eq.~(\ref{eqK4}). In order to obtain explicit analytical expressions, we will resort to a simplifying assumption: in what follows we  assume
that the thermal gradient is sufficiently small to consider only linear terms
in $\Delta T$ in the current Eq.~(\ref{eqK4}), but we shall keep higher order terms
in the finite bias voltage. This is equivalent to write the following
approximation for Eq.~(\ref{eqK4})
\begin{eqnarray}
 I \sim (V/e)G(T,V) + \Delta T L_{12}(T,V) = 0.
\label{eqK8}
\end{eqnarray}
Here, we have defined the coefficient
\begin{eqnarray}
L_{12}(T,V) = -T^{-1}\int d\omega \frac{4\Gamma_{L}\Gamma_{R}}{\Gamma_{L}+\Gamma_{R}} A(\omega,T,V)\omega 
\frac{\partial f_{0}(\omega,T)}{\partial\omega},
 \label{eqK9}
\end{eqnarray}
with $f_{0}(\omega) = \left(e^{\hbar\omega/k_{B}T} + 1 \right)^{-1}$ the Fermi-Dirac distribution. The
nonlinear electrical conductance $G(T,V)$ was already obtained in Eq.~(\ref{eqC7}).
We thus solve for the temperature gradient in Eq.~(\ref{eqK8}):
\begin{eqnarray}
 \left.\Delta T\right|_{I=0} = - (V/e)\frac{G(T,V)}{L_{12}(T,V)}.
\label{eqK10}
\end{eqnarray}
The integral in Eq.~(\ref{eqK7}) is evaluated using the Sommerfeld expansion up to $O(T^{2})$, resulting in
\begin{eqnarray}
 L_{12}(T,V) = T^{-1}\frac{\pi^{2}}{3}\left(k_{B}T \right)^{2}\frac{4\Gamma_{L}\Gamma_{R}}{\Gamma_{L}+\Gamma_{R}}
\frac{\partial A (0,T,V)}{\partial\omega}.
\label{eqK11}
\end{eqnarray}
This expression can be rewritten in the form
\begin{eqnarray}
 L_{12}(T,V) = c_{1}\left(\frac{k_{B}T}{\tilde{\Delta}} \right) - c_{2}\left(\frac{k_{B}T}{\tilde{\Delta}} \right)
\left(\frac{eV}{\tilde{\Delta}} \right)^{2} + O(T^{3},V^{4}),
\label{eqK12}
\end{eqnarray}
where we introduced
\begin{eqnarray}
 c_{1} &=& \frac{8}{9}\pi\zeta\frac{(1-\tilde{u})\tilde{\epsilon}}
{\left[1 + (1-\tilde{u})^{2}\tilde{\epsilon}^{2}\right]^{2}},\nonumber\\
c_{2} &=& \frac{8}{9}\pi\zeta^{2}\tilde{\epsilon}\left[
\frac{2\tilde{u}^{2}(1-\tilde{u})}{\left[1 + (1-\tilde{u})^{2}\tilde{\epsilon}^{2}\right]^{3}}
-\frac{2\tilde{u} + 3\tilde{u}^{2} - 3\tilde{u}^{3}}{6\left[1 + (1-\tilde{u})^{2}\tilde{\epsilon}^{2}\right]^{2}}
\right],
\label{eqK13}
\end{eqnarray}
and the coefficient $\tilde{\epsilon}$ is defined in Eq.~(\ref{eqSPT6}).

\subsection{Thermopower}
\label{subsec:2}
The conditions of applicability of the Mott formula for the Seebeck coefficient of a metal has been discussed by Johnson and Mahan \cite{Jonson.80}.
Horvati\'{c} and Zlati\'{c} \cite{Horvatic.79} used the Mott formula to calculate the thermopower in the asymmetric SIAM.
Mott's formula states that the Seebeck coefficient is related to the energy-dependent scattering relaxation time $\tau(\epsilon)$:
\begin{figure}[t!]
\vspace*{0.8cm}
\sidecaption
\includegraphics[width=7.0cm]{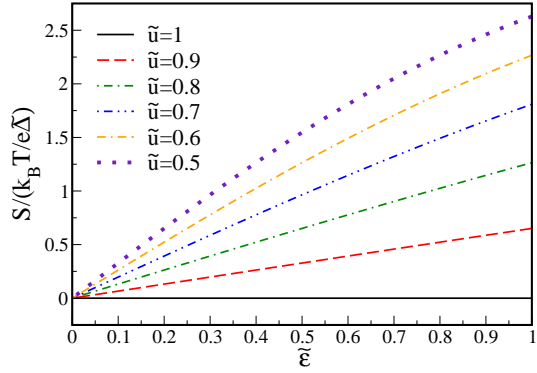}
%
%
\caption{Dependence of the linear response quantity $S/(k_BT)$ in units of $|e|\tilde{\Delta}$ on the renormalized quantities $\tilde{u}$ and $\tilde{\epsilon}$. The thermopower $S$ increases with increasing p-h asymmetry and vanishes when $\tilde{u}\rightarrow 1$ or $\tilde{\epsilon}\rightarrow 0$.}
\label{thermopower}       
\end{figure}
\begin{figure}[t!]
\vspace*{0.8cm}
\sidecaption
\includegraphics[width=7.0cm]{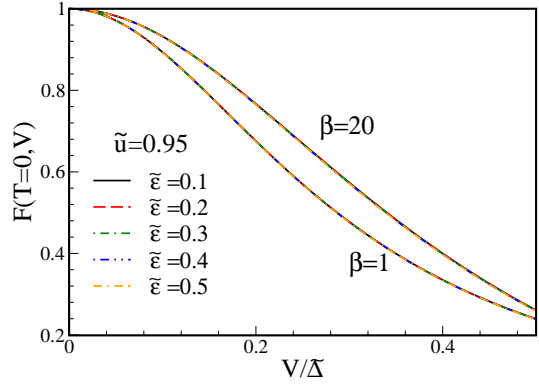}
%
%
\caption{Nonlinear thermopower: Shown is the enhancement factor F defined in Eq.~(\ref{enhanceF}), evaluated at zero temperature for simplicity, versus bias voltage. The behavior near the strong coupling limit $\tilde{u}=0.95$ for various values of the p-h asymmetry $\tilde{\epsilon}$ for symmetric ($\beta=1$) and asymmetric couplings ($\beta=20$). As a result, the nonlinear Seebeck coefficient becomes smaller than the linear response value.}
\label{nonlinearthermopowera}       
\end{figure}
\begin{eqnarray}
S = -\frac{\pi^{2}k_{B}^{2}T}{3|e|}\frac{\partial}{\partial \omega}\left.\ln \tau(\omega)\right. \Big|_{\omega=0}.
\label{eqS1}
\end{eqnarray}
Horvati\'{c} and Zlati\'{c} \cite{Horvatic.79} showed that
the Seebeck coefficient is given by
\begin{eqnarray}
S_{HZ} = \frac{2\pi^{2}k_{B}}{3|e|}\frac{\tilde{E}_{d}/\Delta}{1 + \left(\tilde{E}_{d}/\Delta\right)^{2}}\left(\frac{\tilde{\gamma} k_{B} T}{\Delta} \right)
+ O(T^{3}),
\label{eqS2}
\end{eqnarray}
where $\tilde{E}_{d}$ is  defined as $\tilde{E}_{d} = E_{d} + \left.\Sigma^{R}_{d}(0)\right|_{T=0}$, {\it i.e.}  as the renormalized position of the virtual
bound state, determined in order to satisfy the Friedel sum rule by the condition $\tilde{E}_{d}/\Delta = \cot(\pi\langle n_{d}\rangle)$.
In this equation, the factor $\tilde{\gamma}$, which determines the enhancement of the thermopower, is the inverse of the
quasiparticle Green function renormalization factor, $\tilde{\gamma} = 1 - [\partial \Sigma^{r}_{d}/\partial \omega]_{\omega=0}$.

Here, instead of assuming the applicability of Mott's formula, we apply our previous analysis
for the nonlinear regime to obtain an analytical expression for the Seebeck coefficient.
The temperature gradient as a function of the bias voltage at vanishing electric current is obtained
by substituting Eq.~(\ref{eqK1}) and Eq.~(\ref{eqC1}) into Eq.~(\ref{eqK10}). Differentiating with
respect to the bias voltage, the Seebeck coefficient is obtained up to $O(T^{3},V^{4})$,
\begin{eqnarray}
 S(T,V) &=& \left(\left.\frac{\partial\Delta T}{\partial V}\right|_{I=0} \right)^{-1}\nonumber\\
&=& \frac{2\pi^{2}}{3|e|}\frac{(1-\tilde{u})\tilde{\epsilon}}
{1 + (1-\tilde{u})^{2}\tilde{\epsilon}^{2}}\left(\frac{k_{B}T}{\tilde{\Delta}} \right)
\left[1 + \frac{c_{2}}{c_{1}}
\left(\frac{eV}{\tilde{\Delta}}\right)^{2}\right]
\left\{1 - 2\frac{c_{2}}{c_{1}}\left(\frac{eV}{\tilde{\Delta}}\right)^{2}\right.\nonumber\\
&&\left.
-c_{T}\left(\frac{k_{B}T}{\tilde{\Delta}} \right)^{2}
\left[1 - 2\frac{c_{2}}{c_{1}}\left(\frac{eV}{\tilde{\Delta}}\right)^{2} \right]
+ 3 c_{V}\left(\frac{eV}{\tilde{\Delta}}\right)^{2} - 3 c_{TV}\left(\frac{eV}{\tilde{\Delta}}\right)^{2}
\left(\frac{k_{B}T}{\tilde{\Delta}} \right)^{2}\right.\nonumber\\
&&\left. - 2 c_{VEd}\left(\frac{eV}{\tilde{\Delta}}\right)
+ 3 c_{TVEd}\left(\frac{eV}{\tilde{\Delta}}\right)^{2}\left(\frac{k_{B}T}{\tilde{\Delta}} \right)^{2}\right\}^{-1}.
\label{eq:thermopower}
\end{eqnarray}
At low temperatures and zero bias voltage, 
the full expression Eq.~(\ref{eq:thermopower}) obtained from our theory for the thermopower
can be expressed in the simplified form
\begin{eqnarray}
 S=S(T,V=0) &=& \frac{2\pi^{2}}{3|e|}\frac{k_{B}T}{\tilde{\Delta}}\frac{\tilde{\epsilon}(1-\tilde{u})}{1 + \tilde{\epsilon}^{2}(1-\tilde{u})^{2}}
+ O(T^{3}).
\label{eqS3}
\end{eqnarray}

\begin{figure}[b!]
\vspace*{0.2cm}
\sidecaption
\includegraphics[width=7.1cm]{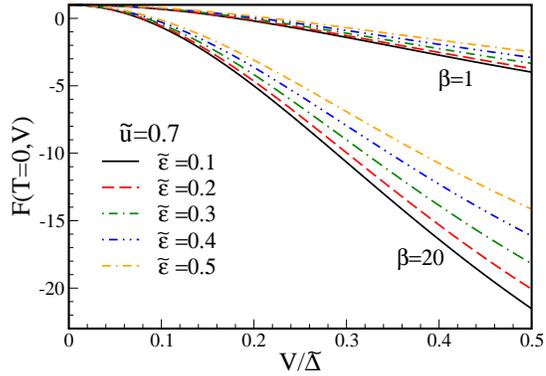}
%
%
\caption{Nonlinear thermopower: The enhancement factor F, defined in Eq.~(\ref{enhanceF}), is evaluated at zero temperature for simplicity. In the regime, where both charge fluctuations and p-h asymmetry
are present, the nonlinear thermopower changes sign as a function of  bias voltage and for sufficient lead-dot asymmetry $\beta$ can become large in magnitude.}
\label{nonlinearthermopowerb}       
\end{figure}
\begin{figure}[t!]
\vspace*{0.8cm}
\sidecaption
\includegraphics[width=7.0cm]{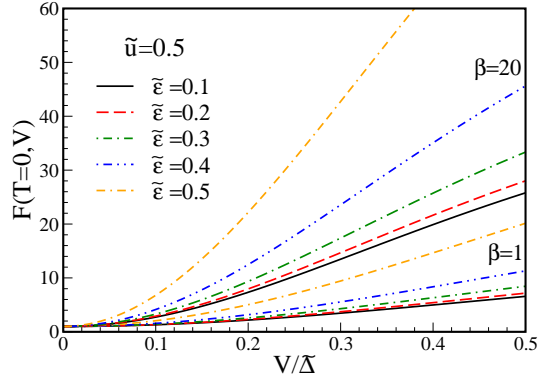}
%
%
\caption{Nonlinear thermopower: Shown is the enhancement factor F defined in Eq.~(\ref{enhanceF}) (here evaluated at zero temperature for simplicity) versus bias voltage for symmetric  lead-to-dot coupling $\beta=1$ and for $\beta=20$ at $\tilde{u}=0.5$, reflecting the presence of charge fluctuations. The enhancement factor turns out to be positive and large implying a huge enhancement of the thermopower in the nonlinear regime.}
\label{nonlinearthermopowerc}       
\end{figure}
It is interesting to compare the thermopower obtained from our superperturbation theory
with the result obtained by  Zlati\'{c} and Horvati\'{c}~\cite{Horvatic.79}.
The renormalized resonance width is $\tilde{\Delta} = \tilde{\chi}_{++}^{-1}\Delta$,
where the spin susceptibility, according to Yamada-Yosida's results, is related to the renormalization factor of
the quasi-particle Green function $\tilde{\gamma} = \tilde{\chi}_{++}$. 
Therefore, in the zero bias voltage limit ({\it i.e.} linear response regime), 
our expression Eq.~(\ref{eq:thermopower}) for the thermopower reduces to Eq.~(\ref{eqS3}) which
is equivalent to the result of Horvati\'{c} and Zlati\'{c}, with the difference that the dependence on the renormalized interaction $\tilde{u}$ is made explicit. As $\lim_{\tilde{u}\rightarrow 1} (1-\tilde{u})\tilde{\epsilon} = 0$. 
Our results
indicate that the 
thermopower decays to zero for a p-h symmetric system,
as well as in the strongly interacting (Kondo) limit, in agreement with the previous theory by Horvati\'{c} and Zlati\'{c} \cite{Horvatic.79}.
This is demonstrated in Fig.~\ref{thermopower}, where  $S/(k_B T/e \tilde{\Delta})$ as a function of p-h asymmetry is
shown for various values of the renormalized interaction strength $\tilde{u}$. 

The nonlinear thermopower, Eq.~(\ref{eq:thermopower}), as a function of  bias voltage  shows a
much richer behavior as compared to the Seebeck coefficient at low temperature.
Comparing Eqs.~(\ref{eq:thermopower}) and (\ref{eqS3}), one can introduce an enhancement factor
via
\begin{equation}
\label{enhanceF}
 F(T,V)=S(T,V)/S,
\end{equation}
such, that $F(T,V=0)=1$. In Fig.~\ref{nonlinearthermopowera}, the behavior of $F(T=0,V)$ near the strong coupling limit is shown for different values of the p-h and lead-dot coupling asymmetry. In parallel to the linear response thermopower, the value of $S(T,V)$ remains small near the strong coupling limit.
The behavior of  $F(T=0,V)$ in the regime, where both charge fluctuations and p-h asymmetry
are present, is shown in Fig.~\ref{nonlinearthermopowerb}.  The enhancement factor in this regime changes sign as a function of  bias voltage and for sufficiently large lead-dot asymmetry $\beta$ can become large in magnitude. Finally, the nonlinear thermopower can become large in the region where p-h asymmetry is present and charge fluctuation are strong (as compared to $\tilde{\epsilon}$) as demonstrated in Fig.\ref{nonlinearthermopowerc}.

\subsection{Thermal Conductance}

We shall now obtain an analytical expression for the thermal conductance, according to 
Eq.~(\ref{eqK7}). We set $K(T,V) = K_{1}(T,V) + K_{2}(T,V)$, corresponding
to the two integral terms in Eq.~(\ref{eqK7}). We calculate $K_{1}(T,V)$, 
by substituting $\partial[f_{L} - f_{R}]/\partial(\Delta T) = -(\omega/T)\partial f_{0}/\partial\omega$ in
the integrand as
follows
\begin{eqnarray}
 K_{1}(T,V) = \frac{4}{3}\frac{\Delta\zeta\hbar}{T}\int_{-\infty}^{+\infty}
d\omega\omega^{2}A(\omega)\left(-\frac{\partial f_{0}}{\partial\omega} \right)
= \frac{4}{3}\Delta\zeta\hbar\frac{\pi^{2}}{3}\frac{k_{B}^{2}T}{\hbar}A(0,V,T),
\label{eqKT1}
\end{eqnarray}
where we have used the Sommerfeld expansion for the Fermi function to evaluate the integral.
\begin{figure}[b!]
\vspace*{0.8cm}
\sidecaption
\includegraphics[width=7.0cm]{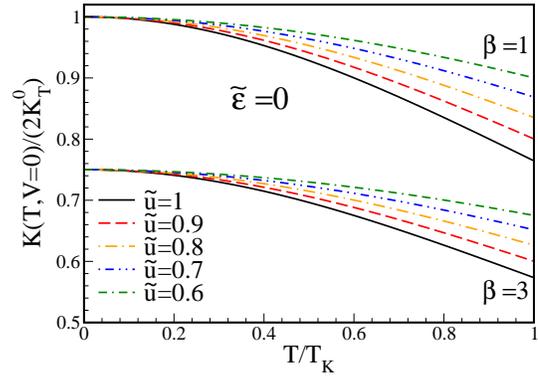}
%
%
\caption{Thermal conductance $K(T,V=0)$ versus temperature in units of twice the quantum of thermal conductance at p-h symmetry $\tilde{\epsilon}=0$ for various values of the renormalized interaction strength $\tilde{u}$.
$K(T,0)$ shows already at the linear response level a strong dependence on the lead-to-dot coupling asymmetry $\beta$ ($k_B T_K=\pi \tilde{\Delta}/4$).}
\label{thermalconductance1}      
\end{figure}
Let us now consider the contribution $K_{2}(T,V)$ arising from the second integral expression in 
Eq.~(\ref{eqK7}),
\begin{eqnarray}
 K_{2}(T,V) &=& S(T,V)\frac{\partial}{\partial V}\left(
\int d\omega \frac{4\Gamma_{L}\Gamma_{R}}{\Gamma_{L} + \Gamma_{R}}\hbar\omega A(\omega,T,V)[f_{L} - f_{R}]
\right)_{\Delta T}.
\label{eqKT2}
\end{eqnarray}
Consistent with the order of approximation of Eq.~(\ref{eqK6}), we expand the difference
of the distribution functions of the leads up to first order in $\Delta T$, but up to third order in 
the voltage gradient,
\begin{eqnarray*}
 f_{L} - f_{R} \sim -\frac{\Delta T}{T}\omega\frac{\partial f_{0}}{\partial\omega} -(eV)\frac{\partial f_{0}}{\partial\omega}
+\frac{1-\beta}{1+\beta}\frac{(eV)^{2}}{2}\frac{\partial^{2} f_{0}}{\partial\omega^{2}}
- (1-\zeta)\frac{(eV)^{3}}{6}\frac{\partial^{3} f_{0}}{\partial\omega^{3}}.
\label{eqKT3}
\end{eqnarray*}
Substituting this expansion into Eq.~(\ref{eqKT2}), we obtain the following expression
\begin{eqnarray}
 K_{2}(T,V) &=& \frac{4}{3}\zeta\Delta\frac{\pi^{2}}{6}(k_{B}T)^{2}S(T,V)
\frac{\partial}{\partial V}\left(2 eV\frac{\partial A(0)}{\partial\omega}
+ \frac{3}{2}(eV)^{2}\frac{1-\beta}{1+\beta}\frac{\partial^{2}A(0)}{\partial\omega^{2}}\right.\nonumber\\
&&\left.+ \frac{2}{3}(1-\zeta)(eV)^{3}\frac{\partial^{3}A(0)}{\partial\omega^{3}} \right),
\label{eqKT4}
\end{eqnarray}
where we  used the Sommerfeld expansion to evaluate the integrals, and the
derivatives of the local spectral function are evaluated at $\omega=0$, but at
finite bias voltage.
It is clear from the prefactor $T^{2}S(T,V) = O(T^{3})$ that $K_{2}(T,V)$
is of $O(T^{3})$ and hence is beyond the order of approximation
$O(T^{2})$ as we assumed from the beginning.  We finally obtain
for the thermal conductance
\begin{eqnarray}
 K(T,V) &=& 2 K_{T}^{0}\frac{4}{3}\zeta\left(1 
+ \frac{\tilde{u}^{2}}{2}\left[\left(\frac{\pi k_{B}T}{\tilde{\Delta}}\right)^{2}
+ \zeta\left(\frac{eV}{\tilde{\Delta}}\right)^{2}\right]\right)\nonumber\\
&\times&\left\{\tilde{\epsilon}^{2}\left[1 
- \tilde{u}\left\{1
 - \frac{1}{3}\left(\frac{\pi k_{B}T}{\tilde{\Delta}}\right)^{2}
-\frac{\zeta}{3}\left(\frac{eV}{\tilde{\Delta}} \right)^{2}
+\frac{2}{3}\zeta\left(\frac{\pi T eV}{\tilde{\Delta}^{2}} \right)^{2}\right\} 
\right]^{2}\right.\nonumber\\
&&\left.+ \left[1 + \frac{\tilde{u}^{2}}{2}\left[\left(\frac{\pi k_{B} T}{\tilde{\Delta}}\right)^{2}
+ \zeta\left(\frac{eV}{\tilde{\Delta}}\right)^{2}\right]\right]^{2}\right\}^{-1}.
\label{eqKT5}
\end{eqnarray}
\begin{figure}[t!]
\vspace*{0.8cm}
\sidecaption
\includegraphics[width=9.0cm]{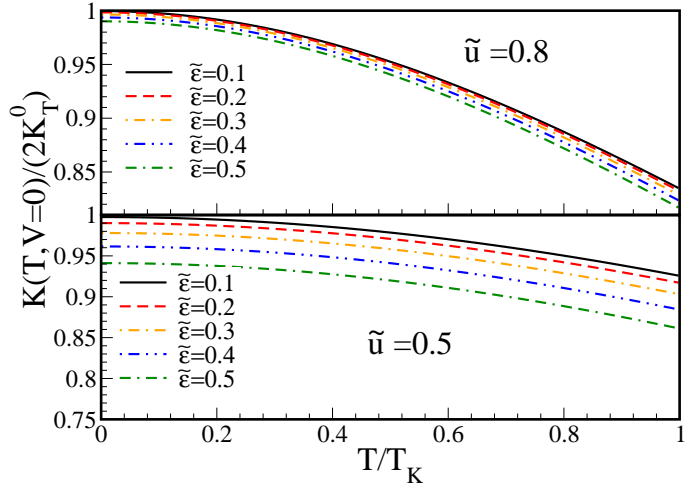}
%
%
\caption{Thermal conductance versus temperature in units of twice the quantum of thermal conductance at $\tilde{u}=0.8$ and $\tilde{u}=0.5$ for various values of the effective p-h symmetry. The dot-lead couplings are equal, $beta=1$. The Kondo-temperature has been defined as $k_B T_K=\pi \tilde{\Delta}/4$.}
\label{thermalconductance2}       
\end{figure}

Here, we have expressed the result in terms of the universal quantum of thermal conductance, 
$K_{T}^{0}\equiv \pi^{2} k_{B}^{2} T/(3h)$. Notice that in the linear response regime, evaluating 
Eq.~(\ref{eqKT5}) at zero bias voltage, we have 
\begin{eqnarray}
K(T,0) = 2 K_{T}^{0}\frac{\frac{4}{3}\zeta(1 + \frac{\tilde{u}^{2}}{2}(\pi k_{B}T/\tilde{\Delta})^{2})}
{\tilde{\epsilon}^{2}\left(1 - \tilde{u}\left[1 - \frac{1}{3}\left(\frac{\pi k_{B}T}{\tilde{\Delta}} \right)^{2} \right] \right)^{2}
+ \left(1 + \frac{\tilde{u}^{2}}{2}(\pi k_{B} T/\tilde{\Delta})^{2} \right)^{2}}.
\label{eqKT6}
\end{eqnarray}
From this later result, we see that as $T\rightarrow 0$, we have
\begin{eqnarray}
\frac{K(T,0)}{K_{T}^{0}} \rightarrow 2\frac{4\zeta/3}{1 + \tilde{\epsilon}^{2}(1 - \tilde{u})^{2}},
\label{eqKT7}
\end{eqnarray}
where the factor of 2 accounts for the two independent spin "channels" of thermal conductance. In the p-hole symmetric case ($\tilde{\epsilon}^{2}=0$) with symmetric contact couplings ($\zeta=3/4$),
the thermal conductance at very low $T$ is just $K_{T} \rightarrow 2K_{T}^{0}$, accounting for two universal quanta of thermal conductance.
In  Fig.~\ref{thermalconductance1}, the thermal conductance $K(T,V=0)$ versus temperature is shown at
the p-h symmetric point. In this case, the zero temperature limit only depends on the lead-to-dot coupling asymmetry. 
Fig.~\ref{thermalconductance2} addresses the dependence of $K(T,V=0)$ on  p-h asymmetry as a function of
temperature.

\subsection{Breakdown of the Wiedemann-Franz law}

\begin{figure}[b!]
\vspace*{0.8cm}
\sidecaption
\includegraphics[width=8.5cm]{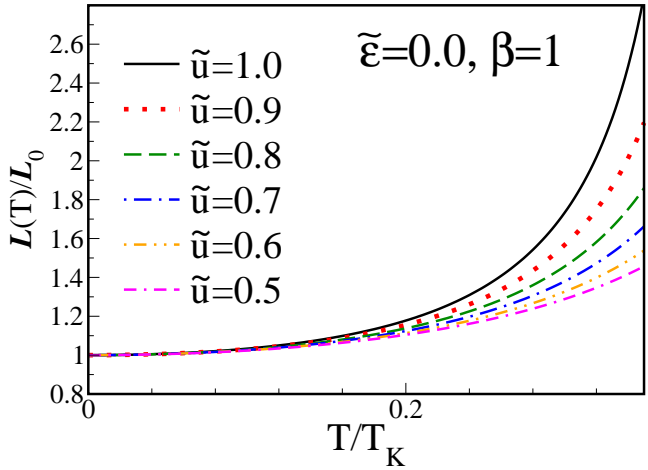}
%
%
\caption{Wiedemann-Franz law: The ratio {${\it L}(T,V=0)/{\it L}_0$} approaches unity in the limit of only elastic scattering, {\it i.e.} at zero temperature. This is demonstrated here for various renormalized interaction strengths at p-h symmetry ($\tilde{\epsilon}=0$).}
\label{WiedemannFranz1}       
\end{figure}
\begin{figure}[t!]
\vspace*{0.8cm}
\sidecaption
\includegraphics[width=8.5cm]{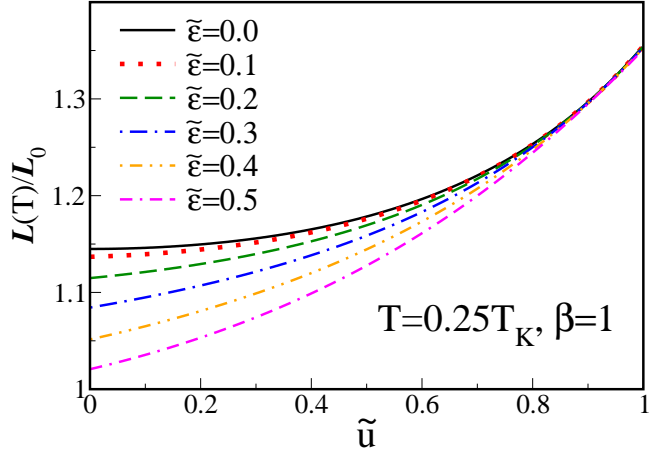}
%
%
\caption{Wiedemann-Franz law: The ratio {${\it L}(T,V=0)/{\it L}_0$} as a function of the renormalized
interaction strength $\tilde{u}$ for various values of the p-h asymmetry $\tilde{\epsilon}$ at 
$T=0.25T_K$. The effect of the p-h asymmetry is irrelevant in the close vicinity of the strong-coupling limit $\tilde{u}\rightarrow 1$.}
\label{WiedemannFranz2}       
\end{figure}

At this level it is interesting to check the range of applicability of the Wiedemann-Franz law.
The Wiedemann-Franz law is not expected to hold when inelastic scattering is present, which happens at finite temperature and is also expected to occur away from thermal equilibrium.

We calculate the Lorenz number ${\it L}$ by taking the ratio of Eq.~(\ref{eqKT5}) for
the thermal conductance, over Eq.~(\ref{eqC1}) for the zero-voltage electrical conductance
\begin{eqnarray}
\label{WF}
{\it L}(T,V) &=& \frac{K(T,V)}{T G(T,V)}\rightarrow \frac{K(T,0)}{T G(T,0)}\\
&=& {\it L}_{0}\frac{1 + (1 - \tilde{u})^{2}\tilde{\epsilon}^{2}}{1 - c_{T}\left(\frac{k_{B}T}{\tilde{\Delta}}\right)^{2}}
\frac{1 + \frac{\tilde{u}^{2}}{2}\left(\frac{\pi k_{B}T}{\tilde{\Delta}} \right)^{2}}{\tilde{\epsilon}^{2}\left(1 - \tilde{u}\left[1 - \frac{1}{3}\left(\frac{\pi k_{B}T}{\tilde{\Delta}} \right)^{2} \right] \right)^{2}
+ \left(1 + \frac{\tilde{u}^{2}}{2}\left(\frac{\pi k_{B} T}{\tilde{\Delta}}\right)^{2} \right)^{2}}.\nonumber
\end{eqnarray}
Here, ${\it L}_{0} = \pi^{2}k_{B}^{2}/(3e^{2})$ is the Lorenz number of a free electron gas. From Eq.~(\ref{WF}),
we easily check that
\begin{eqnarray}
 \lim_{T\rightarrow 0} \frac{{\it L}(T)}{{\it L}_{0}} = 1,
\label{eqKT10}
\end{eqnarray}
which shows that the Wiedemann-Franz law is satisfied in the limit of zero temperature, independent on the strength of the
interactions $\tilde{u}$ and of p-h asymmetry $\tilde{\epsilon}$.

At finite temperatures, inelastic scattering becomes possible and the Wiedemann-Franz ratio starts to deviate from the Lorenz number, as shown in Fig.~\ref{WiedemannFranz1} for various values of $\tilde{u}$ at the p-h symmetric point $\tilde{\epsilon}=0$. The dependence of p-h asymmetry as a function of $\tilde{u}$
is shown for fixed temperature in Fig.~\ref{WiedemannFranz2}.

\begin{figure}[h!]
\vspace*{0.8cm}
\includegraphics[width=\linewidth]{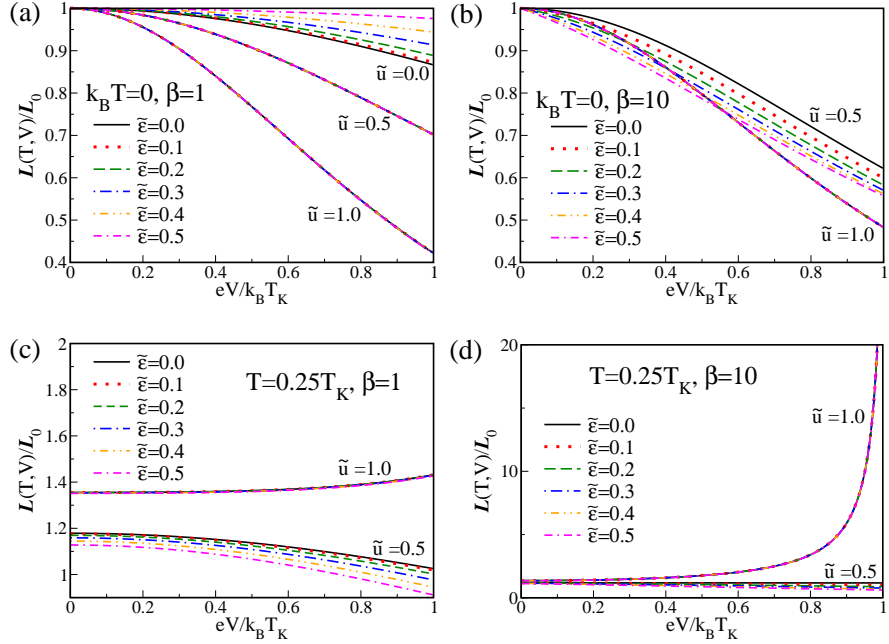}
%
%
\caption{Behavior of the Wiedemann-Franz ratio ${\it L}/{\it L}_0$ defined in Eq.~(\ref{WF}) in the nonlinear regime.for different temperatures and couplings strengths $\beta$ as a function of applied bias voltage $V$. 
(a) The Wiedemann-Franz ratio at zero temperature and equal lead-to-dot coupling of left and right lead, $\beta=1$, for various renormalized interactions $\tilde{u}$ versus p-h asymmetry $\tilde{\epsilon}$. The overall behavior is rather similar and p-h asymmetry only matters at comparatively small interaction strengths. In the limit of vanishing bias voltage, the Wiedemann-Franz law, ${\it L}/{\it L}_0=1$, is recovered.
(b) similar to (a) but with $\beta=10$. 
(c) similar to (a) but at finite temperature $T=0.25T_K$ where $k_B T_K=\pi \tilde{\Delta}/4$. The increase
in ${\it L}/{\it L}_0$ reflects the finite temperature behavior seen at $eV=0$, see Fig.~\ref{WiedemannFranz1}.
(d) similar to (b) but at finite temperature $T=0.25T_K$. The Wiedemann-Franz ratio ${\it L}/{\it L}_0$ displays a strong increase at finite temperature for asymmetric lead-do-couplings in the nonlinear regime.
}
\label{NLWiedemannFranz}       
\end{figure}

In the introduction, we argued that dissipative processes in the nonlinear regime will lead to
deviations from the Wiedemann-Franz law even at zero temperature. That this is indeed the case and that the behavior of the Wiedemann-Franz ratio ${\it L}(T,V)/{\it L}_0$ away from thermal equilibrium is particularly rich, is demonstrated in Fig.~\ref{NLWiedemannFranz} (a)-(d), where ${\it L}/{\it L}_0$ versus bias voltage $eV$ is shown for different regimes and at $T=0$ (in (a) and (b)) and at $T=0.25 T_K$ (in (c) and (d)) for various values of lead-to-dot coupling asymmetry. The most interesting feature is that the deviations from the Wiedemann-Franz law can lead to ${\it L}(T,V)/{\it L}_0 > 1$ as well as ${\it L}(T,V)/{\it L}_0 <1$.
Fig.~\ref{NLWiedemannFranz} (a) shows that for symmetric lead-to-dot coupling, $\beta=1$, p-h asymmetry only matters for very small $\tilde{u}$. Similar results are obtained for asymmetric lead-to-dot coupling, see Fig.~\ref{NLWiedemannFranz} (b). At finite temperature and finite bias voltage the deviations from the Wiedemann-Franz law become more interesting: As seen in  Fig.~\ref{NLWiedemannFranz} (c) and (d),
the strong coupling limit $\tilde{u}\approx 1$ implies a ratio ${\it L}(T,V)/{\it L}_0>0$, that increases strongly with lead-to-dot asymmetry whereas values of $\tilde{u}$ corresponding to intermediate coupling and the presence of charge fluctuations  ${\it L}(T,V)/{\it L}_0$ goes from ${\it L}(T,V)/{\it L}_0>1$ to
${\it L}(T,V)/{\it L}_0<0$ as a function of bias voltage.

\section{Conclusion}

Thermoelectric transport properties of interacting systems beyond the linear-response regime are largely unexplored. This is mainly due to the lack of reliable methods that can treat the interaction problem
away from thermal equilibrium. Yet, there is good reason to believe that a better understanding of nonlinear
thermoelectric transport will help in the search for better thermoelectrics with high figure of merits.\\
To address the nonlinear transport properties
of a strongly correlated system in a well-defined and traceable setting, we here considered the case
of a singleimpurity Anderson model that is driven out of equilibrium by a finite voltage drop and a thermal gradient. Our approach extends a method we recently proposed for the calculation of the nonlinear conductance
of molecular transistors and semiconductor quantum dots (Reference \cite{Munoz.11}) to the calculation of thermoelectric transport coefficients. We reviewed this approach of Ref.~\cite{Munoz.11} focusing on the important issue of current conservation encoded in the distribution function of the dot density of states
in presence of the leads. The entropy production rate in the non-thermal steady-state allowed us to generalize the linear response expressions of the thermoelectric transport coefficients. 
Explicit  expressions for the (linear and nonlinear) thermal conductance
and thermopower are given and the breakdown of the Wiedemann-Franz law in the nonlinear regime is demonstrated. 
As demonstrated, the nonlinear regime of  a quantum dot at intermediate coupling with charge fluctuations
is characterized by an enhanced Seebeck coefficient and a reduced Wiedemann-Franz ratio as compared to the
linear-response quantities.
\begin{acknowledgement}
We thank C.~Bolech, T.~Costi, A.~C.~Hewson, D.~Natelson, J.~Paaske, P.~Ribeiro, G.~Scott and V.~Zlati\'{c} for many stimulating discussions.
E.~M. and S.~K. acknowledge support by the Comisi\'{o}n Nacional de Investigaci\'{o}n Cient\'{i}fica y Tecnol\'{o}gica (CONICYT), grant No. 11100064 
and the German Academic Exchange Service (DAAD) under grant No. 52636698.
\end{acknowledgement}
\section*{Appendix}
\addcontentsline{toc}{section}{Appendix}

\subsection*{A Nonequilibrium Green functions}

\begin{figure}[h!]
\sidecaption
\includegraphics[width=11.0cm]{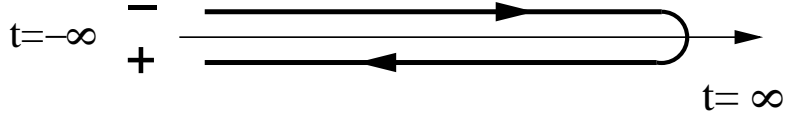}
%
%
\caption{Schwinger-Keldysh contour: The label $\mathbf -$ and $\mathbf +$ denote the time-ordered ($\mathbf -$) and anti-time-ordered ($\mathbf +$) section of the contour.}
\label{KeldyshContour}
\end{figure}

A Green function on the Schwinger-Keldysh contour, see Fig.~\ref{KeldyshContour}, can be defined via
\begin{equation}
{G}(t',t)=-i \langle T{\Psi_H(t')^{}\Psi^\dagger_{H}(t)}\rangle,
\end{equation}
where $T$ is the time ordering operator along the Schwinger-Keldysh contour and
the index of the field operators indicates that they are taken in the  Heisenberg picture.
This gives rise to four different functions, depending on whether $t$ or $t'$ is located on the time-ordered ($\mathbf -$) and anti-time-ordered ($\mathbf +$) section of the contour.
The nonequilibrium Green function can therefore be brought into the form
\begin{eqnarray}
\label{Green}
{\mathbf{G}} = \left[\begin{array}{cc}G^{--} & G^{-+}\\
G^{+-} & G^{++}\end{array} \right],
\end{eqnarray}
where the index
$\mp$ refers to the time-ordered (anti-time-ordered) path in the
Keldysh contour.
The Dyson equation 
\begin{equation}
 {\mathbf G} = {\mathbf g}  + {\mathbf g} \Sigma {\mathbf  G},
\end{equation}
becomes a matrix equation for the selfenergy, where $\bf g$ is the bare Green function.
The components of ${\mathbf{G}}$ are not independent: 
$G^{++}+G^{--}=G^{+-}+G^{-+}$. The retarted and advanced Green functions $G^r$ and $G^a$ are defined as
$G^r=G^{--}-G^{-+}$ and $G^a=G^{--}-G^{+-}$.
Similar relations links the corresponding components of $\Sigma$.

In equilibrium, one finds 
\begin{eqnarray}
G^{-+}_{eq}(\omega) &=& f_{0}(\omega)\left[ G^{r}_{eq}(\omega) - G^{a}_{eq}(\omega)\right],\nonumber\\
G^{+-}_{eq}(\omega) &=& [1 - f_{0}(\omega)] \left[ G^{r}_{eq}(\omega) - G^{a}_{eq}(\omega)\right],\nonumber\\
\Sigma^{-+}_{eq}(\omega) &=& f_{0}(\omega) \left[ \Sigma^{r}_{eq}(\omega) - \Sigma^{a}_{eq}(\omega)\right],\nonumber\\
\Sigma^{+-}_{eq}(\omega) &=& [1-f_{0}(\omega)] \left[ \Sigma^{r}_{eq}(\omega) - \Sigma^{a}_{eq}(\omega)\right],
\end{eqnarray}
where  $f_0=(\exp[\omega/T]+1)^{-1}$ is the Fermi function.

The Dyson equation for the $G^{+-}$ component is given by
\begin{eqnarray}
\label{eq.RHS}
G^{+-}&=&(1+G^r\Sigma^r)g^{+-}(1+\Sigma^a G^a)+G^r \Sigma^{+-} G^a.
\end{eqnarray}
in terms of the retarded and advanced  components of ${\mathbf{G}}$ and
 ${\Sigma}$.

A distribution function can be defined via 
\begin{equation}
G^{+-}={\mathcal{F}}(G^a-G^r),
\end{equation}
We also define a distribution function ${\mathcal{F}}^{*}$ for the self-energy of $G^{+-}$:
\begin{equation}
\label{Sigmadist}
\Sigma^{+-}={\mathcal{F}}^*(\Sigma^a-\Sigma^r).
\end{equation}
The Dyson equation for  $G^a$ (respectively $G^r$) implies
\begin{eqnarray}
\label{relation}
\Sigma^a-\Sigma^r=(G^{-1})^r-(G^{-1})^a,
\end{eqnarray}
or
\begin{eqnarray}
\label{equation}
G^a-G^r=G^r(\Sigma^a-\Sigma^r)G^a,
\end{eqnarray}
and therefore
\begin{eqnarray}
G^{+-}&=&{\mathcal{F}}(G^a-G^r)=G^r{\mathcal{F}}(\Sigma^a-\Sigma^r)G^a.
\end{eqnarray}

 For a general initial state characterized by the distribution function $f_{in}$, defined by $g^{+-}=f_{in}(g^a-g^r)$, the first term of the right hand side of Eq.~(\ref{eq.RHS}) vanishes
\begin{eqnarray}
(1+G^r\Sigma^r)g^{+-}(1+\Sigma^a G^a)&=&f_{in}(1+G^r\Sigma^r)(g^a-g^r)(1+\Sigma^a G^a)\nonumber \\
&=&f_{in}[(1+G^r\Sigma^r)g^a(1+\Sigma^aG^a)-(1+G^r\Sigma^r)g^r(1+\Sigma^aG^a)]\nonumber\\
&=&f_{in}[(1+G^r\Sigma^r)G^a-G^r(1+\Sigma^a G^a)]\nonumber\\
&=&f_{in}(G^a-G^r-G^r\Sigma^a G^a +G^r \Sigma^r G^a)\nonumber\\
&=&f_{in}(G^a-G^r-G^r(\Sigma^a-\Sigma^r) G^a)\nonumber\\
&=&f_{in}(G^a-G^r-(G^a-G^r)) =0,
\end{eqnarray}
where equation (\ref{equation}) was used in the last step.
So,
\begin{eqnarray}
G^{+-}&=&{\mathcal{F}}(G^a-G^r)=G^r{\mathcal{F}}^*(\Sigma^a-\Sigma^r)G^a\\
&=& G^r \Sigma^{+-} G^a,
\end{eqnarray}
where the second line of the last equation is the Dyson equation for the $G^{+-}$ component: $G^{+-}=G^r \Sigma^{+-} G^a$.

Therefore,
\begin{equation}
\label{Sigmadistribution}
\Sigma^{+-}={\mathcal{F}}(\Sigma^a-\Sigma^r),
\end{equation}
and by comparing equation (\ref{Sigmadist}) with (\ref{Sigmadistribution}):
\begin{equation}
\label{resultforF}
{\mathcal{F}}^*(\omega,T,V)={\mathcal{F}}(\omega,T,V).
\end{equation}
Note that the commutativity of ${\mathcal{F}}$ with components of $\hat{G}$ has been assumed. 
The derivation of equation (\ref{relation}) also assumes that we start from the bare Green function $g$ in the infinite past ($t=-\infty$) and turn on the coupling to the leads and the Coulomb interaction $U$ on the dot as the system evolves. Alternatively, one can start with $g$ describing the $u=0$ system that is coupled to the leads. The steady-state properties and therefore relation (\ref{resultforF}) do not depend on the
prescription at $t=-\infty$~\cite{Doyon.06}.

\subsection*{B Renormalized superperturbation theory on the Keldysh contour}
\label{appB}

This appendix summarizes the renormalized superperturbation theory on the Keldysh contour of Ref.~\cite{Munoz.11}.
The term superperturbation theory was used in Ref.~\cite{Hafermann.09} and referred
to a perturbation theory in terms of dual fermions around a fully interacting system solvable via  {\it e.g.} exact diagonalization. In the renormalized superperturbation theory used here~\cite{Munoz.11}, the reference system is based on the work of Yamada and Yoshida \cite{Yosida.70,Yamada.75,Yamada.76} for the symmetric Anderson model is used in the context of renormalized perturbation theory~\cite{Hewson.93}. 

We start from a coherent state representation of the action, {\it i.e.}
\begin{eqnarray}
\hat{c}_{k\lambda\sigma}|k,\lambda,\sigma\rangle &&= c_{k\lambda\sigma}|k,\lambda,\sigma\rangle,\nonumber\\
\langle k,\lambda,\sigma|\hat{c}_{k\lambda\sigma}^{\dagger} &&= c_{k\lambda\sigma}^{*}\langle k,\lambda,\sigma|,\nonumber\\
\hat{d}_{\sigma}|\sigma\rangle &&=  d_{\sigma}|\sigma\rangle,\nonumber\\
\langle\sigma|\hat{d}^{\dagger}_{\sigma} &&= d_{\sigma}^{*}\langle\sigma|,
\label{eqB1}
\end{eqnarray}
where $c_{k\lambda\sigma}$ and $d_{\sigma}$ are Grassmann numbers. The index $\lambda = L, R$ labels the two different leads.

The non-equilibrium "partition function" for the system, in the Keldysh contour (see Fig. 1), is expressed in terms
of a functional integral over time-dependent Grassmann fields, $\hat{\psi}_{k\lambda\sigma}(t) = \left(\begin{array}{c}c_{k\lambda\sigma}^{-}(t)\\c_{k\lambda\sigma}^{+}(t)\end{array} \right)$ and $\hat{\Phi}(t) = \left(\begin{array}{c}d_{\sigma}^{-}(t)\\d_{\sigma}^{+}(t)\end{array} \right)$. Here, the indexes $\pm$ refer to the
time-ordered (-) and anti-time-ordered (+) path along the closed Keldysh contour.
\begin{eqnarray}
Z = \int\mathcal{D}[\hat{\psi}^{\dagger},\hat{\psi}]\mathcal{D}[\hat{\Phi}^{\dagger},\hat{\Phi}]e^{iS[\hat{\psi}^{\dagger},\hat{\psi},\hat{\Phi}^{\dagger},\hat{\Phi}]}.
\label{eqB2}
\end{eqnarray}
The action in Eq.~(\ref{eqB2}) is defined by
\begin{eqnarray}
i S[\hat{\psi}^{\dagger},\hat{\psi},\hat{\Phi}^{\dagger},\hat{\Phi}] &=& i\int_{-\infty}^{+\infty}dt\left\{\sum_{k,\lambda,\sigma}\hat{\psi}_{k\lambda\sigma}^{\dagger}(t)\left(\begin{array}{cc}i\partial_{t}-\epsilon_{k\lambda} &
0\\0 & i\partial_{t}-\epsilon_{k\lambda}\end{array} \right)\hat{\sigma}_{3}\hat{\psi}_{k\lambda\sigma}(t)
\right.\nonumber\\
&&\left.+\sum_{\sigma}\hat{\Phi}^{\dagger}_{\sigma}(t)\left(\begin{array}{cc}i\partial_{t}-E_{d} &
0\\0 & i\partial_{t}-E_{d}\end{array} \right)\hat{\sigma}_{3}\hat{\Phi}_{\sigma}(t) \right.\nonumber\\
&&\left.+\sum_{k,\lambda,\sigma}\left[\hat{\Phi}^{\dagger}_{\sigma}(t)\left(\begin{array}{cc}V_{k\lambda} &
0\\0 & V_{k\lambda}\end{array} \right)\hat{\sigma}_{3}\hat{\psi}_{k\lambda\sigma}(t)\right.\right.\nonumber\\
&&\left.\left. + \hat{\psi}_{k\lambda\sigma}^{\dagger}(t)\left(\begin{array}{cc}V_{k\lambda}^{*} &
0\\0 & V_{k\lambda}^{*}\end{array} \right)\hat{\sigma}_{3}\hat{\Phi}_{\sigma}(t)\right]\right\} + i S_{U}^{int}[\hat{\Phi}^{\dagger},\hat{\Phi}].
\label{eqB3}
\end{eqnarray}
Here, $\hat{\sigma}_{3} = \left(\begin{array}{cc}1 & 0\\0 & -1\end{array} \right)$ is the third Pauli matrix. The Coulomb interaction terms are contained in the action $S_{U}^{int}$ defined by
\begin{eqnarray}
i S_{U}^{int}[\hat{\Phi}^{\dagger},\hat{\Phi}] &=&
i\int_{-\infty}^{+\infty}dt\,\, U\left\{\frac{1}{4}\left[\hat{\Phi}^{\dagger}_{\uparrow}(\hat{\sigma}_{3} + \hat{\sigma}_{0})\hat{\Phi}_{\uparrow} \right]\left[\hat{\Phi}^{\dagger}_{\downarrow}(\hat{\sigma}_{3} + \hat{\sigma}_{0})\hat{\Phi}_{\downarrow} \right]
\right.\nonumber\\
&&\left.-\frac{1}{4}\left[\hat{\Phi}^{\dagger}_{\uparrow}(\hat{\sigma}_{3} - \hat{\sigma}_{0})\hat{\Phi}_{\uparrow} \right]\left[\hat{\Phi}^{\dagger}_{\downarrow}(\hat{\sigma}_{3} - \hat{\sigma}_{0})\hat{\Phi}_{\downarrow} \right]\right.\nonumber\\
&&\left. -
\hat{\Phi}^{\dagger}_{\uparrow}\hat{\sigma}_{3}\hat{\Phi}_{\uparrow}
- \hat{\Phi}^{\dagger}_{\downarrow}\hat{\sigma}_{3}\hat{\Phi}_{\downarrow}\right\},
\label{eqB4}
\end{eqnarray}
where $\hat{\sigma}_{0}$ is the identity matrix.

Since the action in Eq.~(\ref{eqB4}) is Gaussian in the $\hat{\psi}_{k\lambda\sigma}^{\dagger}(t)$, $\hat{\psi}_{k\lambda\sigma}(t)$
Grassmann fields, we integrate those in the partition function Eq.~(\ref{eqB2}) to obtain, in the frequency-space representation,
\begin{eqnarray}
Z = \int\mathcal{D}[\hat{\Phi}^{\dagger}_{\sigma\omega},\hat{\Phi}_{\sigma\omega}]e^{iS[\hat{\Phi}^{\dagger}_{\sigma\omega},\hat{\Phi}_{\sigma\omega}]}.
\label{eqB5}
\end{eqnarray}
In Eq.~(\ref{eqB5}), we defined the effective action as
\begin{eqnarray}
i S[\hat{\Phi}^{\dagger}_{\sigma\omega},\hat{\Phi}_{\sigma\omega}] = i S_{U}[\hat{\Phi}^{\dagger}_{\sigma\omega},\hat{\Phi}_{\sigma\omega}] -
i\int_{-\infty}^{+\infty}\frac{d\omega}{2\pi}\sum_{\sigma}\hat{\Phi}^{\dagger}_{\sigma\omega}E_{d}\hat{\sigma}_{3}\hat{\Phi}_{\sigma\omega},
\label{eqB6}
\end{eqnarray}
where
\begin{eqnarray}
iS_{U}[\hat{\Phi}^{\dagger}_{\sigma\omega},\hat{\Phi}_{\sigma\omega}] = i\int_{-\infty}^{+\infty}\frac{d\omega}{2\pi}
\sum_{\sigma}\hat{\Phi}^{\dagger}_{\sigma\omega}(\omega + i(\Gamma_{L} + \Gamma_{R}))\hat{\sigma}_{3}\hat{\Phi}_{\sigma\omega} +
i S_{U}^{int}[\hat{\Phi}^{\dagger}_{\sigma\omega},\hat{\Phi}_{\sigma\omega}]\nonumber\\
\label{eqB7}
\end{eqnarray}
is the effective action for a p-h symmetric ($E_{d} =0$) and interacting ($U \ne 0$) system, and
\begin{eqnarray}
i\Gamma_{\lambda} = -\sum_{k,\sigma}\frac{|V_{k\lambda}|^{2}}{\omega - \epsilon_{k\lambda} + i\eta^{+}}\,\,\,\,\,\,\,\rm{for}\,\,\, \lambda = L, R
\label{eqB8}
\end{eqnarray}
is the effective tunneling rate the metallic leads, which tends to $\Gamma_{\lambda} \rightarrow \pi\rho_{\lambda}(\omega)|V_{\lambda}|^{2}$ in the limit
of a flat band ($V_{k\lambda} = V_{\lambda}$) of infinite bandwidth, with $\rho_{\lambda}(\omega)=\sum_{k,\sigma}\delta(\omega - \epsilon_{k\lambda})$ the density of states at the $\lambda = L,\,R$ lead.

We bring to bear a super-perturbation scheme~\cite{Rubtsov.08} to treat the term proportional to $E_{d}$ in the effective action Eq.~(\ref{eqB6}),
by using  the p-h symmetric and interacting system described by the effective action $S_{U}[\hat{\Phi}^{\dagger}_{\sigma\omega},\hat{\Phi}_{\sigma\omega}]$ 
in Eq.~(\ref{eqB7}) as a reference system. For that purpose, let us define $\mathbf{g}_{\sigma,\omega}$ as the matrix Green function for the p-h symmetric reference system,
\begin{eqnarray}
{\mathbf{g}}_{\sigma,\omega} = \left[\begin{array}{cc}g^{--}_{\sigma,\omega} & g^{-+}_{\sigma,\omega}\\
g^{+-}_{\sigma,\omega} & g^{++}_{\sigma,\omega}\end{array} \right].
\label{eqB9}
\end{eqnarray}
Let us introduce the dual fermion (Grassmann) fields $\hat{\phi}_{\sigma\omega}=\left(\begin{array}{c}f_{\sigma\omega}^{-}\\f_{\sigma\omega}^{+} \end{array} \right)$ where, as before, the index
$\mp$ refers to the time-ordered (anti-time-ordered) path on the
Keldysh contour. We insert the identity,
\begin{eqnarray}
&&\int\mathcal{D}[\hat{\phi}_{\sigma\omega}^{\dagger},\hat{\phi}_{\sigma\omega}]e^{i\sum_{\sigma}\int_{-\infty}^{+\infty}
\frac{d\omega}{2\pi}\left[\hat{\phi}_{\sigma\omega}^{\dagger}\left(\mathbf{g}_{\sigma\omega}E_{d}\hat{\sigma}_{3}\mathbf{g}_{\sigma\omega}\right)^{-1}\hat{\phi}_{\sigma\omega} - \hat{\phi}_{\sigma\omega}^{\dagger}\mathbf{g}_{\sigma\omega}^{-1}\hat{\Phi}_{\sigma\omega} - \hat{\Phi}_{\sigma\omega}^{\dagger}\mathbf{g}_{\sigma\omega}^{-1}\hat{\phi}_{\sigma\omega}
\right]}\nonumber\\
&=&{\rm{Det}}\left[\left(\mathbf{g}_{\sigma\omega}E_{d}\hat{\sigma}_{3}\mathbf{g}_{\sigma\omega}\right)^{-1}\right]e^{-i\sum_{\sigma}\int_{-\infty}^{+\infty}\frac{d\omega}{2\pi}\hat{\Phi}_{\sigma\omega}^{\dagger}E_{d}\hat{\sigma}_{3}\hat{\Phi}_{\sigma,\omega}},
\label{eqB10}
\end{eqnarray}
into the partition function, to obtain
\begin{eqnarray}
Z &=& Z_{0}\int\mathcal{D}[\hat{\phi}_{\sigma\omega}^{\dagger},
\hat{\phi}_{\sigma\omega}]\mathcal{D}[\hat{\Phi}_{\sigma\omega}^{\dagger},
\hat{\Phi}_{\sigma\omega}] e^{i S_{U}[\hat{\Phi}^{\dagger}_{\sigma\omega},\hat{\Phi}_{\sigma\omega}]}\\
&&\times e^{i\int_{-\infty}^{+\infty}\frac{d\omega}{2\pi}\sum_{\sigma}\hat{\phi}_{\sigma\omega}^{\dagger}(\mathbf{g}_{\sigma\omega}E_{d}\hat{\sigma}_{3})^{-1}
\hat{\phi}_{\sigma\omega} - i\int_{-\infty}^{+\infty}\frac{d\omega}{2\pi}\sum_{\sigma}(\hat{\phi}_{\sigma\omega}^{\dagger}\mathbf{g}_{\sigma\omega}^{-1}\hat{\Phi}_{\sigma\omega} +
\hat{\Phi}_{\sigma\omega}^{\dagger}\mathbf{g}_{\sigma\omega}^{-1}\hat{\phi}_{\sigma\omega} )
}~.\nonumber
\label{eqB11}
\end{eqnarray}
Here, we have defined $Z_{0} = {\rm{Det}}\left[\mathbf{g}_{\sigma\omega}E_{d}\hat{\sigma}_{3}\mathbf{g}_{\sigma\omega} \right]$.
We expand the linear terms in the dual fermion
fields $\hat{\phi}_{\sigma\omega}$ in the action Eq.~(\ref{eqB11}),
and integrate over the fields $\hat{\Phi}_{\sigma\omega}$ to
obtain the effective action
\begin{eqnarray}
i S_{eff}^{f}[\hat{\phi}^{\dagger}_{\sigma\omega},\hat{\phi}_{\sigma\omega}] = i\int_{-\infty}^{+\infty}\frac{d\omega}{2\pi}\sum_{\sigma}\hat{\phi}_{\sigma\omega}^{\dagger}[\mathbf{G}_{\sigma\omega}^{f(0)}]^{-1}\hat{\phi}_{\sigma\omega} + i S_{int}^{f}[\hat{\phi}_{\sigma\omega}^{\dagger},\hat{\phi}_{\sigma\omega}],
\label{eqB12}
\end{eqnarray}
where the bare dual fermion Green function is defined by
\begin{eqnarray}
\mathbf{G}_{\sigma\omega}^{f(0)} = \left[-\mathbf{g}_{\sigma\omega}^{-1} +\left(\mathbf{g}_{\sigma\omega}E_{d}\hat{\sigma}_{3}\mathbf{g}_{\sigma\omega}\right)^{-1}\right]^{-1}
= -\mathbf{g}_{\sigma,\omega}\left(\mathbf{g}_{\sigma,\omega} -
E_{d}^{-1}\hat{\sigma}_{3}\right)^{-1}
\mathbf{g}_{\sigma,\omega}.
\label{eqB13}
\end{eqnarray}
One obtains a direct relation between the dual fermion Green
function $\mathbf{G}_{\sigma\omega}^{f}$ and the Green function
for localized states in the dot, by noticing that the partition
function can be written in two equivalent ways \cite{Rubtsov.08}, Eq.~(\ref{eqB11})
and Eqs.(\ref{eqB6},\ref{eqB7}),
\begin{eqnarray}
Z &=& \int\mathcal{D}[\hat{\Phi}_{\sigma\omega}^{\dagger},\hat{\Phi}_{\sigma\omega}] e^{i S_{U}[\hat{\Phi}_{\sigma\omega}^{\dagger},\hat{\Phi}_{\sigma\omega}] - i\int_{-\infty}^{+\infty}\frac{d\omega}{2\pi}
\sum_{\sigma}\hat{\Phi}_{\sigma\omega}^{\dagger}E_{d}\hat{\sigma}_{3}\hat{\Phi}_{\sigma\omega}}\\
 &=& Z_{0}\int\mathcal{D}[\hat{\phi}_{\sigma\omega}^{\dagger},
\hat{\phi}_{\sigma\omega}]\mathcal{D}[\hat{\Phi}_{\sigma\omega}^{\dagger},
\hat{\Phi}_{\sigma\omega}] e^{i S_{U}[\hat{\Phi}^{\dagger}_{\sigma\omega},\hat{\Phi}_{\sigma\omega}]}\nonumber\\
&&\times e^{i\int_{-\infty}^{+\infty}\frac{d\omega}{2\pi}\sum_{\sigma}\hat{\phi}_{\sigma\omega}^{\dagger}(\mathbf{g}_{\sigma\omega}E_{d}\hat{\sigma}_{3})^{-1}
\hat{\phi}_{\sigma\omega} - i\int_{-\infty}^{+\infty}\frac{d\omega}{2\pi}\sum_{\sigma}(\hat{\phi}_{\sigma\omega}^{\dagger}\mathbf{g}_{\sigma\omega}^{-1}\hat{\Phi}_{\sigma\omega} +
\hat{\Phi}_{\sigma\omega}^{\dagger}\mathbf{g}_{\sigma\omega}^{-1}\hat{\phi}_{\sigma\omega} )
}.\nonumber
\label{eqB14}
\end{eqnarray}
We define the matrix $\hat{D}_{\sigma\omega} = E_{d}\hat{\sigma}_{3}$,
and take the functional derivative on both sides of Eq~(\ref{eqB14})
to obtain,
\begin{eqnarray}
\left.\frac{1}{Z}\frac{\delta Z}{\delta \hat{D}_{\sigma\omega}}\right
|_{\hat{D}_{\sigma\omega}=E_{d}\hat{\sigma}_{3}}
&=& -i\langle \hat{\Phi}_{\sigma\omega}^{\dagger}\otimes\hat{\Phi}_{\sigma\omega}\rangle\\ 
&=& \frac{1}{Z_{0}}
\left.\frac{\delta Z_{0}}{\delta \hat{D}_{\sigma\omega}}\right
|_{\hat{D}_{\sigma\omega}=E_{d}\hat{\sigma}_{3}}
 + \left(\mathbf{g}_{\sigma\omega}E_{d}\hat{\sigma}_{3}\right)^{-1}i\langle \hat{\phi}_{\sigma\omega}^{\dagger}\otimes\hat{\phi}_{\sigma\omega}\rangle
\left(E_{d}\hat{\sigma}_{3}\mathbf{g}_{\sigma\omega}\right)^{-1}.\nonumber
\label{eqB15}
\end{eqnarray}
By noticing that $\mathbf{G}_{\sigma\omega} = -i\langle \hat{\Phi}_{\sigma\omega}\otimes \hat{\Phi}_{\sigma\omega}^{\dagger}\rangle$,
$\mathbf{G}_{\sigma\omega}^{f} = -i\langle \hat{\phi}_{\sigma\omega}\otimes \hat{\phi}_{\sigma\omega}^{\dagger}\rangle$, and
the simple result
$Z_{0}^{-1}\left.\frac{\delta Z_{0}}{\delta \hat{D}_{\sigma\omega}}\right
|_{\hat{D}_{\sigma\omega}=E_{d}\hat{\sigma}_{3}} = E_{d}^{-1}\hat{\sigma}_{3}$, we obtain from Eq~(\ref{eqB15}) the relation,
\begin{eqnarray}
\mathbf{G}_{\sigma,\omega} = -E_{d}^{-1}\hat{\sigma}_{3} +
\left(\mathbf{g}_{\sigma,\omega}E_{d}\hat{\sigma}_{3}\right)^{-1}
\mathbf{G}_{\sigma,\omega}^{f}\left(E_{d}\hat{\sigma}_{3}\mathbf{g}_{\sigma,\omega}\right)^{-1}.
\label{eqB16}
\end{eqnarray}
In this equation, $\mathbf{G}_{\sigma,\omega}$ is the Green function
matrix for the interacting ($U\ne 0$) asymmetric ($E_{d}\ne 0$)
Anderson model, while $\hat{\sigma}_{3}$ is the third Pauli matrix. In contrast,
$\mathbf{g}_{\sigma,\omega}$ is the
Green function for the interacting ($U\ne 0$) and symmetric ($E_{d}=0$) Anderson
model. Finally, $\mathbf{G}_{\sigma,\omega}^{f}$ is the dual
fermion matrix Green function, obtained from the solution of the matrix Dyson
equation
\begin{eqnarray}
\mathbf{G}_{\sigma,\omega}^{f} = \mathbf{G}_{\sigma,\omega}^{f(0)}
+ \mathbf{G}_{\sigma,\omega}^{f(0)}\Sigma_{\sigma,\omega}^{f}
\mathbf{G}_{\sigma,\omega}^{f} \simeq \mathbf{G}_{\sigma,\omega}^{f(0)}
+ \mathbf{G}_{\sigma,\omega}^{f(0)}\Sigma_{\sigma,\omega}^{f}
\mathbf{G}_{\sigma,\omega}^{f(0)}.
\label{eqB17}
\end{eqnarray}
Here, the bare dual fermion Green function is defined by Eq.~(\ref{eqB13}).
The dual fermion selfenergy $\Sigma_{\sigma,\omega}^{f}$ is obtained from the renormalized four-point vertex of the reference system. This is vital for obtaining the correct behavior in the stong-coupling limit $\tilde{u}\rightarrow 1$ and for  ensuring current conservation in this scheme.

From our calculation,
the retarded component of the self-energy is given by the expression
\begin{eqnarray}
\Sigma_{E_{d}}^{r} 
&=& (1-\tilde{\chi}_{++})\omega + E_{d}
- \frac{E_{d}}{\tilde{\chi}_{++}}\left(\frac{U}{\pi\Delta} \right)
\left\{1 - \frac{\tilde{\chi}_{++}^{2}}{3}\left[\left(\frac{\pi T}{\Delta} \right)^{2}
+\zeta\left(\frac{eV}{\Delta} \right)^{2} \right]\right.\\
&&\left.+ \frac{2\zeta}{3}\tilde{\chi}_{++}^{4}\left(\frac{\pi T e V}{\Delta^{2}} \right)^{2} \right\}
-i\frac{\Delta}{2}\left(\frac{U}{\pi\Delta}\right)^{2}\left[\left(\frac{\omega}{\Delta}
\right)^{2}+\left(\frac{\pi T}{\Delta} \right)^{2} +
\zeta\left(\frac{eV}{\Delta}\right)^{2} \right].\nonumber
 \label{eqB18}
\end{eqnarray}
Here, the parameter
\begin{eqnarray}
\zeta = 3\frac{\beta}{(1 + \beta)^{2}},
 \label{eqB19}
\end{eqnarray}
for $\beta = \Gamma_{L}/\Gamma_{R}$, is a measure of the asymmetry in the coupling to the leads. 
The retarded local ({\it i.e.} on the dot) Green function  is therefore
\begin{eqnarray}
\label{RGreenfunction}
 G_{\sigma,\omega}^{r} &=& (\omega + i\Delta - \Sigma_{E_{d}}^{r})^{-1}\nonumber\\
&=&\tilde{\chi}_{++}^{-1}\left(\omega - \tilde{E}_{d} + i\tilde{\Delta} + \tilde{E}_{d}\tilde{u}\left\{1 - \frac{1}{3}\left[\left(\frac{\pi T}{\tilde{\Delta}} \right)^{2}+\zeta\left(\frac{eV}{\tilde{\Delta}} 
\right)^{2}\right]
+ \frac{2\zeta}{3}\right.\right.\nonumber\\
&&\left.\left.\times\left(\frac{\pi T e V}{\tilde{\Delta}^{2}} \right)^{2}
\right\}
+i\frac{\tilde{\Delta}}{2}\tilde{u}^{2}\left[\left(\frac{\omega}{\tilde{\Delta}} \right)^{2} + \left(\frac{\pi T}{\tilde{\Delta}} \right)^{2}
+ \zeta\left(\frac{eV}{\tilde{\Delta}} \right)^{2} \right]\right)^{-1},
\end{eqnarray}
where the renormalized parameters are given by the expressions:
$\tilde{u} = \tilde{\chi}_{++}^{-1}U/(\pi\Delta)$,
which represents the renormalized Coulomb interaction,
$\tilde{\epsilon} \equiv E_{d}/\Delta$,
representing the p-h asymmetry relative to the width of the resonance, and
$ \tilde{\Delta} \equiv \tilde{\chi}_{++}^{-1}\Delta$,
being the renormalized width of the quasiparticle resonance.
The renormalization factor for the quasiparticle Green function is given by $\tilde{\chi}_{++}^{-1}$, with the spin susceptibility
given by the perturbation theory result obtained by Yamada and Yosida \cite{Yosida.70,Yamada.75,Yamada.76},
\begin{eqnarray}
\tilde{\chi}_{++}= 1 + (3 - \pi^{2}/4)\left(U/\pi\Delta\right)^{2}.
\label{eqB20}
\end{eqnarray}
The corresponding renormalized local spectral function $A(\omega,T,V) =-(1/\pi)Im\,G_{\sigma,\omega}^{r}$ within our renormalized superperturbation theory is given by 
\begin{eqnarray}
A(\omega,T,V) &=&\frac{\tilde{\chi}_{++}^{-1}}{\pi\tilde{\Delta}}\left(1 + \frac{1}{2}\tilde{u}^{2}\left[\left(\frac{\omega}{\tilde{\Delta}} \right)^{2} + \left(\frac{\pi T}{\tilde{\Delta}} \right)^{2}
+ \zeta\left(\frac{eV}{\tilde{\Delta}} \right)^{2} \right]\right)\left\{\left(\frac{\omega}{\tilde{\Delta}} - \frac{\tilde{E}_{d}}{\tilde{\Delta}}\right.\right.\nonumber\\
&&\left.\left. + \frac{\tilde{E}_{d}}{\tilde{\Delta}}\tilde{u}
\left\{1 - \frac{1}{3}\left[\left(\frac{\pi T}{\tilde{\Delta}} \right)^{2}+\zeta\left(\frac{eV}{\tilde{\Delta}} \right)^{2}
\right]+ \frac{2\zeta}{3}\left(\frac{\pi T e V}{\tilde{\Delta}^{2}} \right)^{2}\right\}  \right)^{2}\right.\nonumber\\
&&\left.+ \left(1 + \frac{1}{2}\tilde{u}^{2}\left[\left(\frac{\omega}{\tilde{\Delta}} \right)^{2} + \left(\frac{\pi T}{\tilde{\Delta}} \right)^{2}
+ \zeta\left(\frac{eV}{\tilde{\Delta}} \right)^{2} \right]  \right)^{2}
\right\}^{-1}.
\label{B21}
\end{eqnarray}

As the potential scattering term is marginally irrelevant, our perturbative approach in $E_d$ is expected to work well~\cite{Munoz.11}. Furthermore, the dual fermions have the appealing property that for $E_d\gg 1$,
the dual fermion Green function is simply proportional to the  Green function of the reference system:
\begin{eqnarray}
\mathbf{G}_{\sigma\omega}^{f(0)}
= -\mathbf{g}_{\sigma,\omega}\left(\mathbf{g}_{\sigma,\omega} -
E_{d}^{-1}\hat{\sigma}_{3}\right)^{-1}
\mathbf{g}_{\sigma,\omega}\approx -\mathbf{g}_{\sigma,\omega} ~\mbox{for}~E_d \gg 1.
\label{eqB22}
\end{eqnarray}
For $\tilde{u}=0$, Eq.~(\ref{RGreenfunction}) reduces to
\begin{equation}
G^r_{\sigma,\omega}=\frac{1}{\omega-E_d+i\Delta(\omega)},
\label{eqB23}
\end{equation}
reproducing the non-interacting limit exactly~\cite{Munoz.11}.

\end{document}